\documentclass[12pt,a4paper,final]{iopart}
\usepackage{iopams} 
\expandafter\let\csname equation*\endcsname\relax
\expandafter\let\csname endequation*\endcsname\relax
\usepackage{amsmath}
\usepackage{graphicx}
\usepackage{enumerate}
\usepackage{fancyhdr}
\usepackage[toc,page]{appendix}
\usepackage[breaklinks=true,colorlinks=true,linkcolor=blue,urlcolor=blue,citecolor=blue]{hyperref}
\graphicspath{{figuresNJP/}}

\newcommand{\+}{\textrm{+}}
\newcommand{\mst}{\textrm{-}}
\newcommand{\changefont}{%
    \fontsize{9}{11}\selectfont
}
\begin{document}

\title[Quantum state estimation when qubits are lost]{Quantum state estimation when qubits are lost: A no-data-left-behind approach}
\author{Brian P. Williams and Pavel Lougovski}
\address{Quantum Information Science Group, Oak Ridge National Laboratory, Oak Ridge, Tennessee USA 37831}
\ead{williamsbp@ornl.gov; lougovskip@ornl.gov}

\begin{abstract}
We present an approach to Bayesian mean estimation of quantum states using hyperspherical parametrization and an experiment-specific likelihood which allows utilization of all available data, even when qubits are lost. With this method, we report the first closed-form Bayesian mean estimate for the ideal single qubit. Due to computational constraints, we utilize numerical sampling to determine the Bayesian mean estimate for a photonic two-qubit experiment in which our novel analysis reduces burdens associated with experimental asymmetries and inefficiencies. This method can be applied to quantum states of any dimension and experimental complexity.\footnote{\scriptsize This manuscript has been authored by UT-Battelle, LLC under Contract No. DE-AC05-00OR22725 with the U.S. Department of Energy.  The United States Government retains and the publisher, by accepting the article for publication, acknowledges that the United States Government retains a non-exclusive, paid-up, irrevocable, world-wide license to publish or reproduce the published form of this manuscript, or allow others to do so, for United States Government purposes.  The Department of Energy will provide public access to these results of federally sponsored research in accordance with the DOE Public Access Plan(
\href{http://energy.gov/downloads/doe-public-access-plan}{http://energy.gov/downloads/doe-public-access-plan}). }\end{abstract}

\noindent{\it \small Keywords}:\small quantum state estimation, qubit, Bayesian mean estimation\\
\normalsize
\tableofcontents

\pagestyle{fancy}

\section{Introduction}
The problem of estimating the state of a quantum system from observed measurement outcomes has been around since the dawn of quantum mechanics. In recent years, driven by technological advances in building and controlling quantum systems, this question has received a renewed interest especially in the context of quantum information processing (QIP). Since QIP relies on one's ability to prepare arbitrary multi-qubit quantum states, verifying experimentally that a quantum state has been indeed prepared acceptably close (by some metric) to a target is of paramount importance. As a result, methods built on classical statistical parameter estimation procedures have been adopted in order to satisfy a demand for quantum state characterization tools.

Arguably the most widespread quantum state estimation approach relies on classical maximum likelihood estimation (MLE) technique. Pioneered in~\cite{Hradil}, it offers a great simplicity in numerical implementation but suffers from several dangerous flaws. Most prominently, MLE is prone to rank-deficient estimates of a density matrix from a {\it finite} number of measurement samples~\cite{blume2010optimal} which in turn implies that probability of observing certain states is zero -- a statement that is only valid in the limit of {\it infinite} number of observations. Also, MLE does not provide a straightforward way to place error bars on an estimated quantum state. 

Fortunately, there is an alternative parameter estimation technique called the Bayesian mean estimate (BME) that can be applied to quantum state characterization~\cite{blume2010optimal,jones1991principles,Granade2016} and is free of the the aforementioned shortcomings. In addition, BME minimizes the mean square error(MSE)~\cite{blume2010optimal,jaynes2003probability} i.e. the average square difference between the parameter and its estimate. Thus, BME offers a more accurate estimate of a quantum state. But, BME poses an implementation challenge. It computes a posterior distribution over quantum states for given measurement data by using Bayes' rule which in turn requires one to calculate the probability of the data by integrating over the manifold of all physical quantum states. While it may be possible to carry out the multi-dimensional integration analytically, ultimate evaluation still may be computationally prohibitive. Thus, numerical routines that use Monte Carlo (MC) methods are applied in order to sample from the posterior distribution. There is a trade-off between the speed and accuracy of BME depending on which MC algorithm is used. To our knowledge two types of MC algorithms were proposed for quantum state BME so far. The first one is the Metropolis-Hastings~\cite{MetropolisHastings} (MH) algorithm--an example of Markov Chain MC (MCMC)--was adopted in~\cite{blume2010optimal}. The second one is sequential MC (SMC) ~\cite{del2006sequential}--an importance sampling based algorithm--recently used for adaptive quantum state tomography~\cite{adaptive}. The MH algorithm is known for its ability to reproduce probability distributions very accurately at the expense of slow convergence. On the other hand, the SMC algorithm is fast but may converge to a sample that does not faithfully represent the distribution of interest.

When to apply  or the BME depends on many factors. For instance, for a small measurement data set and a large number of unknown parameters--a typical situation for multi-qubit systems--the BME is superior to  as we demonstrate in Section~\ref{performance} of this paper. But perhaps even more crucially, the applicability of the BME approach depends on the choice of the form and parametrization of the likelihood function. The experimental likelihood most used in application is a simple multinomial that connects the observed data set directly with the quantum probabilities using Born's rule \cite{Hradil}. This approach assumes the data set, observed measurement outcomes, result directly and only from the quantum state and unitary operations. This is not always the case, since the measurement apparatus often introduces operation bias (not always unitary) and inefficiencies that modify the probability of an experimental observation. Previously, James et al. \cite{James01}  accounted for bias in qubit operations in their two-photon tomography method using MLE. More recent MLE works such as Gate Set Tomography \cite{stark2014self,blume2013robust} assume nothing about the qubit state or operations, gates, other than their dimension. However, previous methods stop short of full accounting for non-unitary operations such as qubit loss. Thus, experimentalists may find themselves applying normalizing constants to account for deficiencies in the defined likelihood. These normalizing constants require preliminary experiments to obtain, and the method for obtaining these constants is often not well defined nor reported.

In this paper we develop a BME-based quantum state reconstruction method that utilizes the slice sampling~\cite{sliceSampling} (SS) algorithm which has the accuracy of the MH algorithm but demonstrates faster convergence~\cite{NealConv} and is more resilient for a numerical implementation. We show that by using the hyperspherical parameterization of the manifold of density matrices the BME of a state of a single qubit can be computed analytically by using a uniform prior. For a two-qubit system, in a situation when individual qubits may be lost during the measurement process, we apply SS algorithm to the same parameterization and an experiment-specific likelihood demonstrating a computationally stable and efficient way of sampling from the posterior distribution over the density matrices. We compare the resulting BME estimates to the corresponding MLE estimates as a function of the number of measurements and observe the superiority of the BME method, especially in the limit of small sample sizes. 

We begin this paper with a quick outline of our method in Section~\ref{Sec:outline}. We derive a closed BME for the ideal single qubit experiment in Section~\ref{Sec:singlequbit}. This approachable example illustrates our method and contrasts it with traditional MLE methods. It may also inspire further research into closed-form BME solutions of higher dimensional quantum systems. Next, in Section~\ref{Sec:twoqubits}, we derive a likelihood for a finite data two-photon experiment where detector inefficiencies and experimental asymmetries are taken into account. Utilization of this approach results in the real world benefit of eliminating the need to perform preliminary experiments to determine normalization constants.  Subsequently, in Section~\ref{performance} we simulate a multitude of two-qubit photon experiments generating data sets from which we compare the performance of various MLE and BME approaches. Lastly, we apply our estimation to a real world two-photon experiment in Section~\ref{Sec:Experimental}. In the Appendices, we describe a common MLE approach using a traditional likelihood, we detail numerical procedures for sampling density matrices from the true state distribution using slice sampling, and describe the optimization method used in likelihood maximization.

\section{Approach Outline}\label{Sec:outline}
\begin{figure}[t]
\centering
\includegraphics[width=\textwidth]{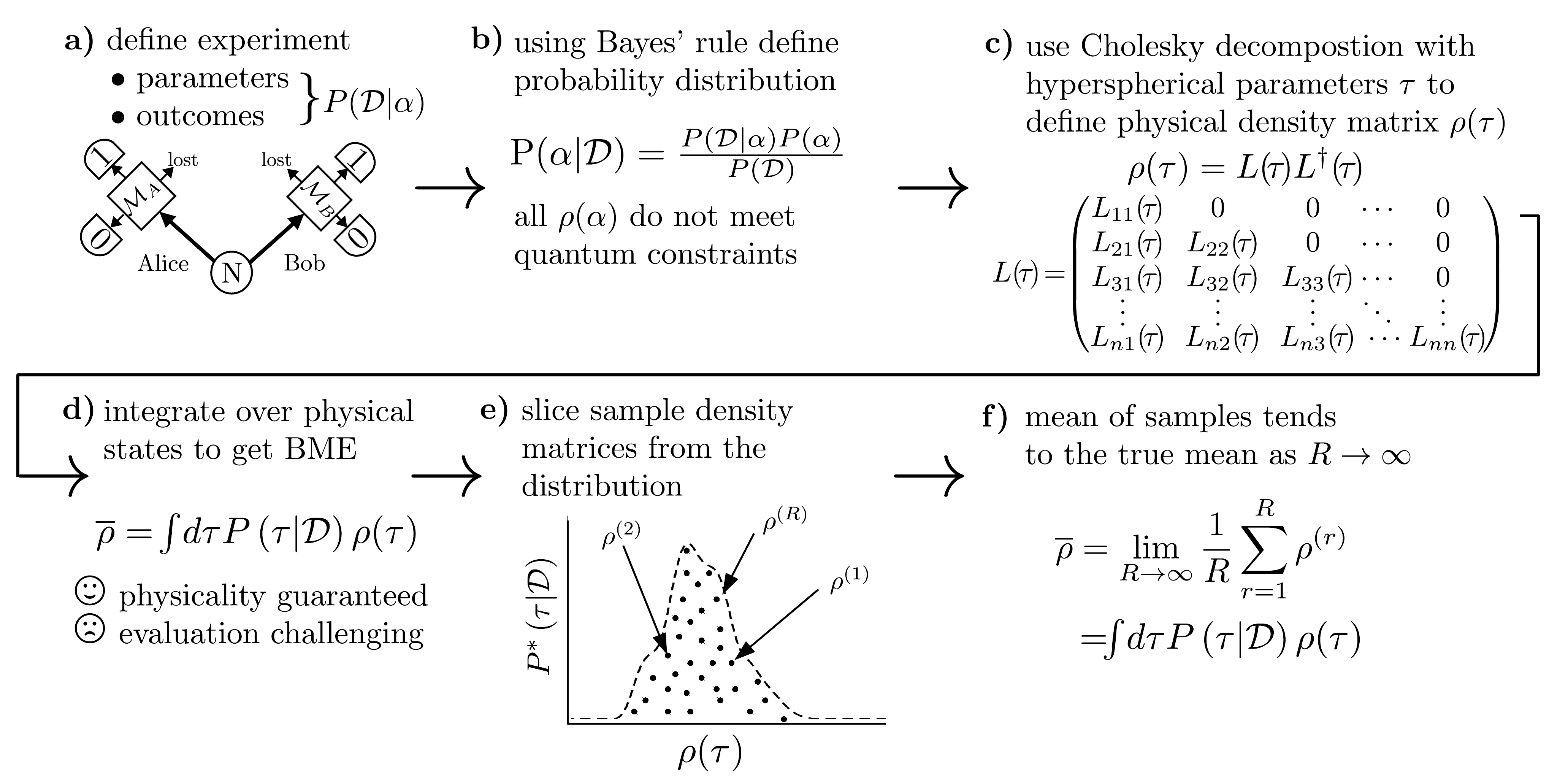}
\caption{Our Bayesian mean estimation of density matrices is outlined above. a) We define our experiment by specifying a likelihood function $\!P\left(\mathcal{D}|\alpha\right)$ of data $\mathcal{D}$ given parameters $\alpha$ and b) express a corresponding posterior distribution of the parameters $\alpha$ which define the density matrix given data using Bayes' rule. c) We parametrize the density matrix such that any choice of parameters in a specified range leads to a valid physical state. d) We represent our posterior distribution using these new parameters; now the BME of the density matrix can be formally written down as an integral using a Haar-invariant measure. Unfortunately, the integral's analytical solution is typically computationally intensive to evaluate. e) Thus, we use computationally efficient numerical slice sampling to make samples of $\rho$ from the postirior distribution $\!P\left(\tau|\mathcal{D}\right)$. f) As $R$$\rightarrow$$\infty$ we tend to the true mean $\overline{\rho}$.\label{outline}}\end{figure}

The components of our quantum state estimation pipeline are outlined in Fig. \ref{outline}. First, we define a model of our experiment by enumerating all the possible outcomes. This enumeration allows us to specify an experiment-specific likelihood $P(\mathcal{D}|\alpha)$, the probability of observing a specific data set $\mathcal{D}$ given the experimental parameters $\alpha=\{\alpha_{1},\cdots,\alpha_{N}\}$. In our case parameters $\alpha$ are elements of a density matrix $\rho$ representing the quantum state to be estimated. Bayes' rule,
\begin{equation}P(\alpha|\mathcal{D})=\frac{P(\mathcal{D}|\alpha)P(\alpha)}{P(\mathcal{D})}\end{equation}
then allows us to express $P(\alpha|\mathcal{D})$, a posterior distribution (PD) for the variables $\alpha$, given an observed data set $\mathcal{D}$ and a prior probability distribution $P(\alpha)$. 

Next, the BME for a specific parameter $\alpha_i$ given a data set $\mathcal{D}$ is
\begin{equation}\overline{\alpha_i}=\int  d\alpha P(\alpha|\mathcal{D})\times \alpha_i \textrm{.}\end{equation}

We expand our analysis to quantum systems by assuming that parameters $\alpha$ are entries of a density matrix $\rho$ ($\rho_{ij}=\alpha_{k}$) describing a valid quantum state (i.e. $\rho\ge 0$, $\rho=\rho^{\dagger}$, $\textrm{Tr}(\rho)=1$). Therefore, $\alpha$'s are not independent as we must enforce quantum constraints. To achieve this in a computationally tractable fashion, instead of using the Cartesian parameterization given by $\alpha$'s we parametrize a density matrix $\rho$ utilizing a Cholesky decomposition (see panel {\bf c} in Fig.(\ref{outline})) and hyperspherical parameters as suggested by Daboul \cite{daboul1967conditions}. We abbreviate this parametrization with $\tau$ to distinguish it from the Cartesian parametrization $\alpha$.

Next, in order to compute the BME estimator, we need to select a prior probability distribution over density matrices $P(\tau)\equiv P(\rho(\tau))$ and an integration measure over the set of all physical quantum states $d\tau$ such that $d\mu(\rho(\tau))=P(\tau)d\tau$ is a valid probability measure i.e. $\int d\mu(\rho(\tau)) = 1$. We use a non-informative prior $P(\tau) = \textrm{const}$ and derive the integration measure $d\tau$ induced by the Riemanian metric $g_{ij}$ computed from the Euclidian length element between density matrices $(ds)^{2} = \textrm{Tr}\left(d\rho(\tau)\cdot d\rho^{\dagger}(\tau)\right)$~\cite{fyodorov2005introduction}. This choice of the integration measure guarantees Haar invariance of the probability measure over the set of density matrices. Thus, the probability of a state $\rho(\tau)$ is invariant under an arbitrary unitary rotation $U$ i.e. $P(\rho(\tau))=P(U\rho(\tau) U^{\dagger})$. Then the BME of an unknown quantum state reads,
\begin{equation}\overline{\rho}=\int d\tau P(\tau|\mathcal{D})\times \rho(\tau).\end{equation}

The latter expression for the BME can, in principle, be evaluated analytically. However, in practice it almost surely requires computational resources and (or) time constraints that prohibit analytical evaluation. In this case, an estimate can be obtained using numerical sampling from the posterior distribution $P(\tau|\mathcal{D})$. For example, in later sections we utilize numerical slice sampling \cite{sliceSampling} to arrive at approximate estimates for a two-photon experiment.

\section{An example: Bayesian mean estimation of an ideal single qubit}\label{Sec:singlequbit}
Consider an ideal single-qubit experiment. In this experiment we can reliably and repetitively prepare a qubit in an unknown state $\rho$ and measure the value of any desired observable $M$ without err or qubit loss. If $M$ is a two outcome POVM defined by operators $M_i$ with $i\in\{0,1\}$ , $M_0 + M_1 = I$  then the respective probabilities $p_i$ to observe outcome $i$ are determined by the unknown quantum state via $p_i = \textrm{Tr}\left(\rho\cdot M_i\right)$. To fully describe a single qubit we need to measure a set of informationally complete POVMs which will fully define the density matrix. For concreteness let us consider a case when the qubit is represented by the polarization degree of freedom of a single photon. In this case a complete state description can be achieved by estimating the probability of observing one of two orthogonal outcomes in the rectilinear basis ($Z$, horizontal ($h$) and vertical ($v$) polarization), the diagonal basis ($X$, diagonal ($d$) and anti-diagonal ($a$) polarization), or the circular basis ($Y$, left ($l$) and right ($r$) circular polarization). The likelihood of observing a data set $\mathcal{D}$ from these measurements given we know the probabilities of each outcome exactly is
\begin{equation}P(\mathcal{D}|\alpha)=p_h^{c_h} (1\mst p_h)^{c_v} p_d^{c_d}(1\mst p_d)^{c_a} p_l^{c_l} (1\mst p_l)^{c_r}\label{likelihood}\end{equation}
where $\alpha=\{p_h,p_d,p_l\}$, $\mathcal{D}=\{c_h,c_v,c_d,c_a,c_l,c_r\}$, we have enforced the single basis requirement that the sum of orthogonal probabilities is unity, $p_{h,d,l}+p_{v,a,r}=1$. Using Bayes rule, the distribution for $\alpha$ given $\mathcal{D}$ is
\begin{equation}P(\alpha|\mathcal{D})=\frac{P(\mathcal{D}|\alpha)P(\alpha)}{\int d\alpha P(\mathcal{D}|\alpha)P(\alpha)}\end{equation}
which has no quantum constraints, i.e. associated density matrices may not be physical. A physical density matrix $\rho$ for the single-qubit must fulfill constraints
\begin{eqnarray}\textrm{Tr}\left(\rho\right)=1 \quad &&\textrm{probabilities sum to 1} \nonumber \\
\left\langle \phi \right| \rho \left|\phi\right\rangle \geq 0 \quad &&\textrm{positive semi-definite} \nonumber \\
\rho=\rho^\dagger \quad &&\textrm{hermitian} \textrm{.}\nonumber \end{eqnarray}
These can all be fulfilled by parametrizing the density matrix as suggested by Daboul \cite{daboul1967conditions}. For the single-qubit the parametrized matrix is 
\begin{equation}\rho(\tau)\!=\!\left(\!\begin{array}{cc}
    \cos^2\left(u\right) & \frac{1}{2}\cos\left(\theta\right)\sin\left(2u\right)e^{i\phi} \\
    \frac{1}{2}\cos\left(\theta\right)\sin\left(2u\right)e^{-i\phi} & \sin^2\left(u\right)
    \end{array}\!\right)\label{ideal_rho}\end{equation}
where parameter ranges $\tau=\{u,\theta,\phi\}$, $u\in[0,\frac{\pi}{2}]$, $\theta\in[0,\frac{\pi}{2}]$, and $\phi\in[0,2\pi]$ ensure there is no state redundancy, states having multiple representations. This matrix heeds all quantum constraints for any values of the parameters. The parameters $\alpha$ in terms of the new parameters $\tau$ are
\begin{align}p_h(\tau) &= \cos^2\left(u\right)\label{pH}\\
p_d(\tau)&=\frac{1}{2}+\frac{1}{2}\sin\left(2u\right)\cos\left(\theta\right)\cos\left(\phi\right)\label{pV}\\
p_l(\tau)&=\frac{1}{2}+\frac{1}{2}\sin\left(2u\right)\cos\left(\theta\right)\sin\left(\phi\right)\textrm{.}\label{pL}\end{align}
This results in the new likelihood
\begin{equation}P(\mathcal{D}|\tau)=p_h(\tau)^{c_h} (1\mst p_h(\tau))^{c_v} p_d(\tau)^{c_d}(1\mst p_d(\tau))^{c_a} p_l(\tau)^{c_l} (1\mst p_l(\tau))^{c_r}\label{tau_likelihood}\textrm{.}\end{equation}
To complete the new description we must define a new integration measure in $\tau$ space. Our original probability space has an infinitesimal length element $(ds)^2=(dp_h)^2+(dp_d)^2+(dp_r)^2$. The measure in this case is reduced to the volume element in Cartesian coordinates $d\alpha=dp_h dp_d dp_l$. This space can be considered a``cube" that includes both physical and unphysical states. Within this cube, the new space is a sphere containing only and all physical density matrices. The  length element in this space is \cite{fyodorov2005introduction}
\begin{equation}(ds)^2=\textrm{Tr}\left(d\rho \cdot d\rho^\dagger\right)=\sum\limits_{i,j}\textrm{Tr}\left(\frac{\partial \rho}{\partial \tau_i}\cdot\frac{\partial \rho}{\partial \tau_j}\right)d\tau_i d\tau_j\label{measure}\end{equation}
where $\tau_i\in\{u,\theta,\phi\}$. The new measure, the infinitesimal volume, is
\begin{equation}d\tau = d\tau_0\; d\tau_1 ..d\tau_m \textrm{Det}\sqrt{g} \label{dTau}\end{equation}
where 
\begin{equation}g_{i j}=\textrm{Tr}\left(\frac{\partial \rho}{\partial \tau_i}\cdot\frac{\partial \rho}{\partial \tau_j}\right)\textrm{.}\label{gIJ}\end{equation}
The integration measure in the ideal single qubit experiment is
\begin{equation}d\tau = du\; d\theta\; d\phi\;\frac{\textrm{sin}^3\left(2u\right)\textrm{sin}\left(2\theta\right)}{2\sqrt{2}} \textrm{.}\nonumber\end{equation}
As described earlier, this measure is Haar invariant.

We will also consider how this parametrization relates to the Pauli operators
\begin{equation}\sigma_z=
\left(\!\begin{array}{cc}
1 & 0\\
0 & -1 \\ \end{array}\!\right) \quad \sigma_x=
\left(\!\begin{array}{cc}
0 & 1\\
1 & 0 \\ \end{array}\!\right)\quad \sigma_y=
\left(\!\begin{array}{cc}
0 & -i\\
i & 0 \\ \end{array}\!\right)
\end{equation}
and their expectations
\begin{align}z&=\textrm{Tr}\left(\sigma_z\cdot\rho\right)=\cos(2u)\label{pauliZ}\\
x&=\textrm{Tr}\left(\sigma_x\cdot\rho\right)=\sin(2u)\cos(\theta)\cos(\phi)\label{pauliX}\\
y&=\textrm{Tr}\left(\sigma_y\cdot\rho\right)=\sin(2u)\cos(\theta)\sin(\phi)\label{pauliY}\textrm{.}
\end{align}

With our likelihood defined, one estimation technique is to approximate the true  distribution utilizing Laplace's method \cite{mackay2003information}, also known as the saddle-point approximation. This is a multivariate Gaussian centered on  the MLE defined by $k$ parameters. This MLE is found by simultaneously solving $k$ equations of the form
\begin{equation}\frac{\partial P(\mathcal{D}|\tau)}{\partial \tau_i}=0\end{equation}
and verifying this point represents the global maximum. The uncertainty in the parameters can be captured utilizing the covariance matrix which we estimate as 
\begin{equation}A_{ij}=\left.-\frac{\partial^2 \log\left(P(\mathcal{D}|\tau)\right)}{\partial \tau_i \partial \tau_j}\right|_{\tau=\tau_{\textrm{ml}}}\textrm{.}\end{equation}
The approximate distribution is then
\begin{equation}P(\mathcal{D}|\tau)\approx\sqrt{\frac{(2\pi)^k}{\det\left(\mathbf{A}\right)}}e^{-\frac{1}{2}\left(\mathbf{\tau}-\mathbf{\tau}_{mle}\right)^T \cdot A \cdot\left(\mathbf{\tau}-\mathbf{\tau}_{mle}\right)}\end{equation}
where $\mathbf{\tau}$ is a column vector.

For the ideal single-qubit we find unbounded MLE
\begin{equation}u_{\textrm{uml}}=\frac{\arccos\left(z_f\right)}{2}\quad\quad
\theta_{\textrm{uml}}=\arccos\left(\sqrt{\frac{x_f^2+y_f^2}{1-z_f^2}}\right)\quad\quad
\phi_{\textrm{uml}}=\arctan\left(x_f,y_f\right)\label{unbounded}\end{equation}
where $z_f$, $x_f$, and $y_f$ are the frequency based linear inversion estimates (LIE) of the Pauli operator expectations
\begin{equation}
z_f=\frac{c_h-c_v}{c_h+c_v}\quad\quad\quad x_f=\frac{c_d-c_a}{c_d+c_a}\quad\quad\quad y_f=\frac{c_l-c_r}{c_l+c_r}\textrm{.}\label{unbounded2}
\end{equation}
When $x_f^2+y_f^2+z_f^2\leq1$ these LIE are the correct MLE. However the parameter set given in Eq. \ref{unbounded} and \ref{unbounded2} is undefined for unphysical states, when $x_f^2+y_f^2+z_f^2>1$. When this is the case, the MLE is found on the boundary of the Bloch sphere due to the concavity of the likelihood given by Eq. \ref{tau_likelihood}. This point is not necessarily the one with smallest Euclidean distance to the unbounded MLE. Determination of the boundary MLE is accomplished by setting $\theta=0$, restricting us to the boundary, and maximizing the parametrized likelihood Eq. \ref{tau_likelihood} over the parameter ranges $u\in\left[0,\frac{\pi}{2}\right]$ and $\phi\in\left[0,2\pi\right]$. Next, we derive a closed-form BME which always results in a quantum bound obedient estimate.

To calculate the BME for the single qubit density matrix we first evaluate the normalizing constant
\begin{equation}P(\mathcal{D})=\int d\tau P(\mathcal{D}|\tau)P(\tau)\end{equation}
and then estimate our mean density matrix
\vspace{-5pt}
\begin{align}\overline{\rho}&=\frac{1}{P(\mathcal{D})}\int d\tau P(\mathcal{D}|\tau)P(\tau)\times \rho\;(\tau)\nonumber\\
&=\frac{1}{P(\mathcal{D})}\int du\; d\theta\; d\phi\; \frac{\textrm{sin}^3\left(2u\right)\textrm{sin}\left(2\theta\right)}{2\sqrt{2}}\left(\cos^2 u\right)^{c_h}\left(\sin^2 u\right)^{c_v} \nonumber \\
&\quad\times \left(\frac{1}{2}+\frac{1}{2}\sin\left(2u\right)\cos\left(\theta\right)\cos\left(\phi\right)\right)^{c_d}\left(\frac{1}{2}-\frac{1}{2}\sin\left(2u\right)\cos\left(\theta\right)\cos\left(\phi\right)\right)^{c_a}\nonumber \\
&\quad\times \left(\frac{1}{2}+\frac{1}{2}\sin\left(2u\right)\cos\left(\theta\right)\sin\left(\phi\right)\right)^{c_l}\left(\frac{1}{2}-\frac{1}{2}\sin\left(2u\right)\cos\left(\theta\right)\sin\left(\phi\right)\right)^{c_r}\nonumber \\
&\quad\times \left(\!\begin{array}{cc}
    \cos^2\left(u\right) & \frac{1}{2}\cos\left(\theta\right)\sin\left(2u\right)e^{i\phi} \\
    \frac{1}{2}\cos\left(\theta\right)\sin\left(2u\right)e^{-i\phi} & \sin^2\left(u\right)
    \end{array}\!\right)\nonumber \textrm{.}
\end{align}
Using the binomial theorem we can rewrite this as
\small
\begin{align}&\overline{\rho}\;=\frac{1}{P(\mathcal{D})}\int_0^{\pi/2}\!\!\!\!\!du\int_0^{\pi/2}\!\!\!\!\!d\theta\int_0^{2\pi}\!\!\!\!\!d\phi\;  \frac{8\; \textrm{sin}^3\left(u\right)\textrm{cos}^3\left(u\right)\sin\left(\theta\right)\cos\left(\theta\right)}{\sqrt{2}}\left(\cos^2 u\right)^{c_h}\left(\sin^2 u\right)^{c_v} \nonumber \\
&\times\!\!\sum_{k_d=0}^{c_d}\!\!\binom{c_d}{k_d} 2^{-k_d} \!\left(\sin(u)\cos(u)\cos(\theta)\cos(\phi)\right)^{c_d-k_d}\sum_{k_a=0}^{c_a}\!\!\binom{c_a}{k_a} 2^{\mst k_a} \!\left(\mst\sin(u)\cos(u)\cos(\theta)\cos(\phi)\right)^{c_a-k_a}\nonumber \\
&\times\!\!\sum_{k_l=0}^{c_l}\!\!\binom{c_l}{k_l} 2^{-k_r} \!\left(\sin(u)\cos(u)\cos(\theta)\sin(\phi)\right)^{c_l-k_l}\sum_{k_r=0}^{c_r}\!\!\binom{c_r}{k_r} 2^{\mst k_l} \!\left(\mst\sin(u)\cos(u)\cos(\theta)\sin(\phi)\right)^{c_r-k_r}\nonumber \\
&\quad\times \left(\!\begin{array}{cc}
    \cos^2\!\left(u\right) & \frac{1}{2}\cos\!\left(\theta\right)\sin\!\left(2u\right)e^{i\phi} \\
    \frac{1}{2}\cos\!\left(\theta\right)\sin\!\left(2u\right)e^{-i\phi} & \sin^2\!\left(u\right)
    \end{array}\!\right)\nonumber \textrm{.}
\end{align}
\normalsize
The integral over $u$ has solution
\begin{equation}\int_0^{\pi/2} du \sin^x\!u \cos^y\!u =\frac{1}{2}\;\textrm{Beta}\left(\frac{1+x}{2},\frac{1+y}{2}\right)\end{equation}
and similar for $\theta$. The integral over $\phi$ can be shown to be 
\begin{equation}\int_0^{2 \pi} d\phi\; \sin^x \phi \cos^y \phi= \frac{\left(1+(\textrm{-}1)^x\right)\left(1+(\textrm{-}1)^y\right)}{2}\;\textrm{Beta}\left(\frac{1+x}{2},\frac{1+y}{2}\right)\end{equation}
which is zero for odd $x$ or $y$. To ease representation of the solutions, define
\scriptsize
\begin{align}&F_{u_0,u_1,\theta_0,\theta_1,\phi_0,\phi_1}\nonumber \\
&=\int_0^{\pi/2}\!\!\!\!\!du\int_0^{\pi/2}\!\!\!\!\!d\theta\int_0^{2\pi}\!\!\!\!\!d\phi\;  \frac{8\; \textrm{sin}^3\left(u\right)\textrm{cos}^3\left(u\right)\sin\left(\theta\right)\cos\left(\theta\right)}{\sqrt{2}} \left(\cos^2 u\right)^{c_h}\left(\sin^2 u\right)^{c_v} \nonumber \\
&\quad\times\!\!\sum_{k_d=0}^{c_d}\!\!\binom{c_d}{k_d} 2^{-k_d} \!\left(\sin(u)\cos(u)\cos(\theta)\cos(\phi)\right)^{c_d-k_d}
\sum_{k_a=0}^{c_a}\!\!\binom{n_d\mst c_d}{k_a} 2^{-k_a} \!\left(-\sin(u)\cos(u)\cos(\theta)\cos(\phi)\right)^{c_a-k_a}\nonumber \\
&\quad\times\!\!\sum_{k_l=0}^{c_l}\!\!\binom{c_r}{k_r} 2^{-k_r} \!\left(\sin(u)\cos(u)\cos(\theta)\sin(\phi)\right)^{c_l-k_l} \sum_{k_r=0}^{c_r}\!\!\binom{n_c\mst c_r}{k_l} 2^{-k_l} \!\left(-\sin(u)\cos(u)\cos(\theta)\sin(\phi)\right)^{c_r-k_r}\nonumber \\
&\quad\times \cos^{u_0}(u)\sin^{u_1}(u)\cos^{\theta_0}(\theta)\sin^{\theta_1}(\theta)\cos^{\phi_0}(\phi)\sin^{\phi_1}(\phi) \nonumber \\
&=\sum_{k_d=0}^{c_d}\sum_{k_d=0}^{c_d}\sum_{k_l=0}^{c_l}\sum_{k_r=0}^{c_r}\binom{c_d}{k_d}\binom{c_a}{k_a}\binom{c_l}{k_l}\binom{c_r}{k_r} 2^{-k_d-k_a-k_l-k_r}(\mst 1)^{c_a+c_r-k_a-k_r}\!\left(\!1\!+\!(\mst 1)^{n_c \mst k_l \mst k_r \+ \phi_0}\right)\left(\!1\!+\!(\mst 1)^{n_d \mst k_d \mst k_a \+ \phi_1}\right)\nonumber \\
&\quad\times\textrm{Beta}\left(\frac{4 \+ 2\;c_h + n_d +n_c - k_d - k_a - k_l - k_r + u_0}{2},\frac{4 + 2\;c_v + n_d + n_c - k_d - k_a - k_l - k_r + u_1}{2}\right)\nonumber \\
&\quad\times\textrm{Beta}\left(\frac{2+\theta_0}{2},\frac{1 + n_d + n_c - kd - ka - k_l - k_r + \theta_1}{2}\right)\;\textrm{Beta}\left(\frac{1 + n_c - k_l - k_r + \phi_0}{2},\frac{1 + n_d - k_d - k_a + \phi_1}{2}\right)\nonumber\textrm{.}
\end{align}
\normalsize
The BME for our ideal single-qubit is
\begin{equation}\overline{\rho}= \frac{1}{F_{0,0,0,0,0,0}}
\left(\!\begin{array}{cc}
F_{2,0,0,0,0,0} & F_{1,1,0,1,0,1}\+ i F_{1,1,0,1,1,0}\\
F_{1,1,0,1,0,1}\mst i F_{1,1,0,1,1,0} & F_{0,2,0,0,0,0} \\\end{array}\!\right)\textrm{.}\nonumber
\end{equation}
This is the best possible estimation of the ideal single-qubit given a set of data $\mathcal{D}$ and a uniform prior. 
\begin{figure}[b]
\centering
\includegraphics[scale=0.5]{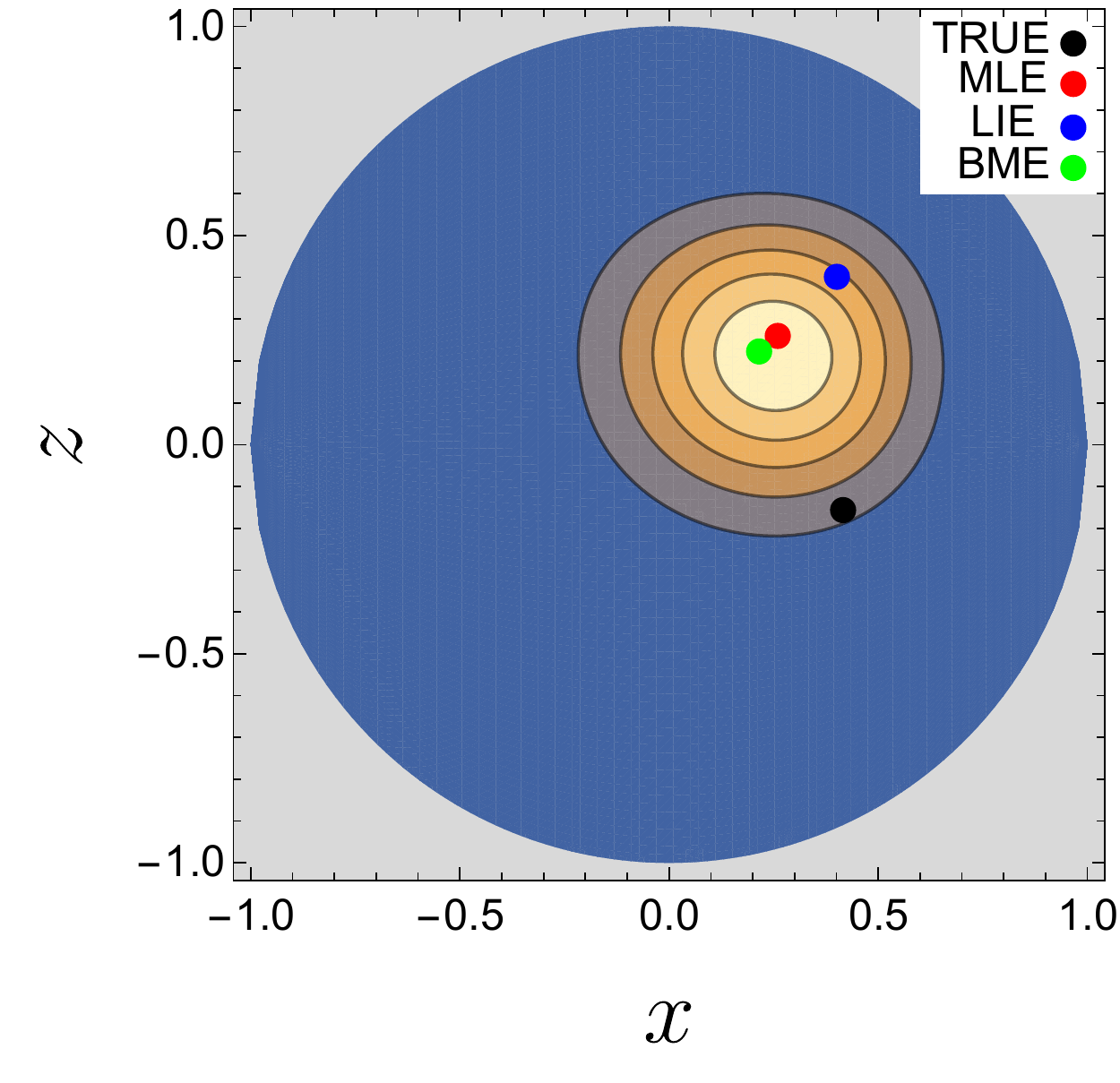}
\includegraphics[scale=0.5]{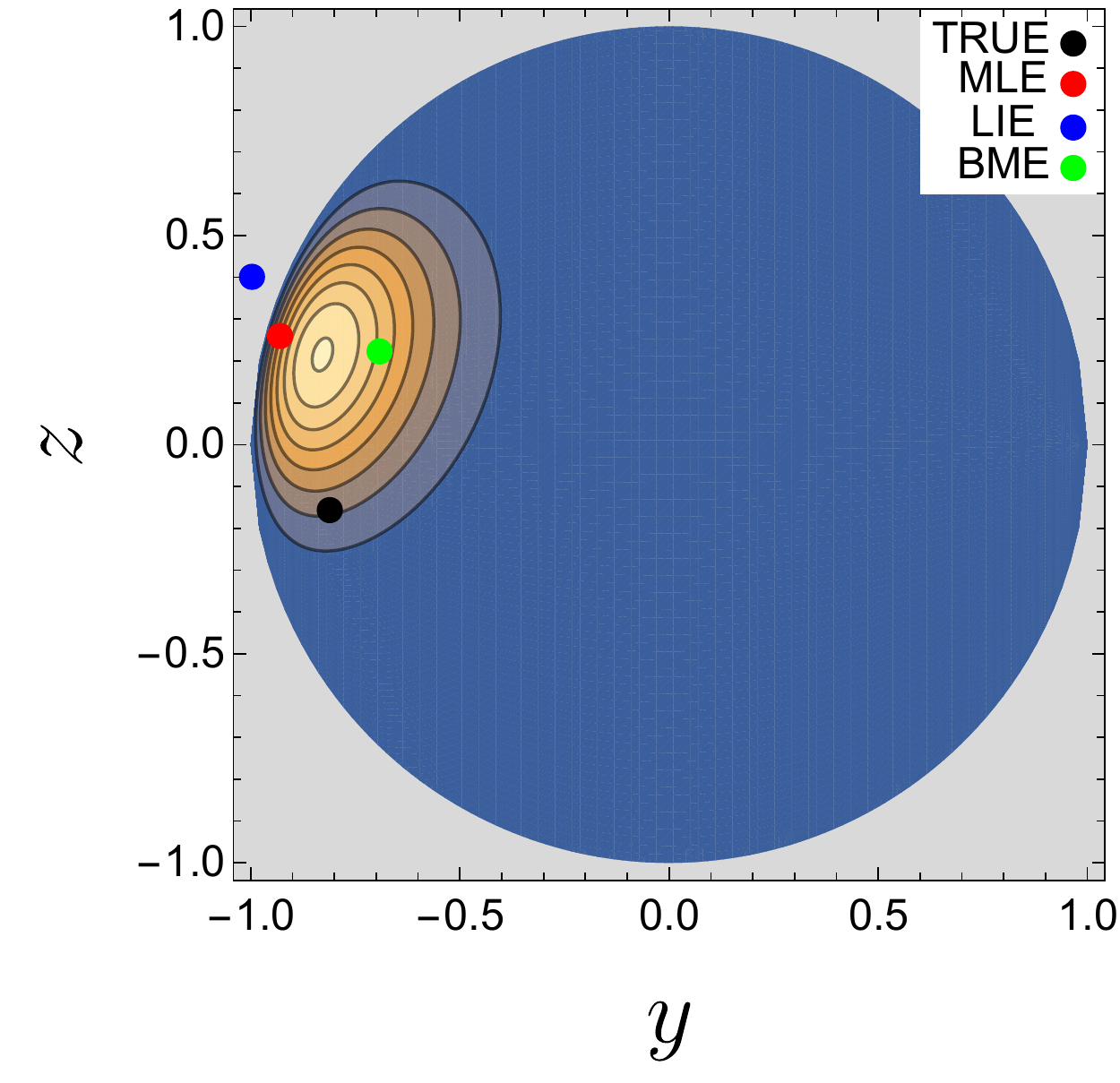}
\includegraphics[scale=0.5]{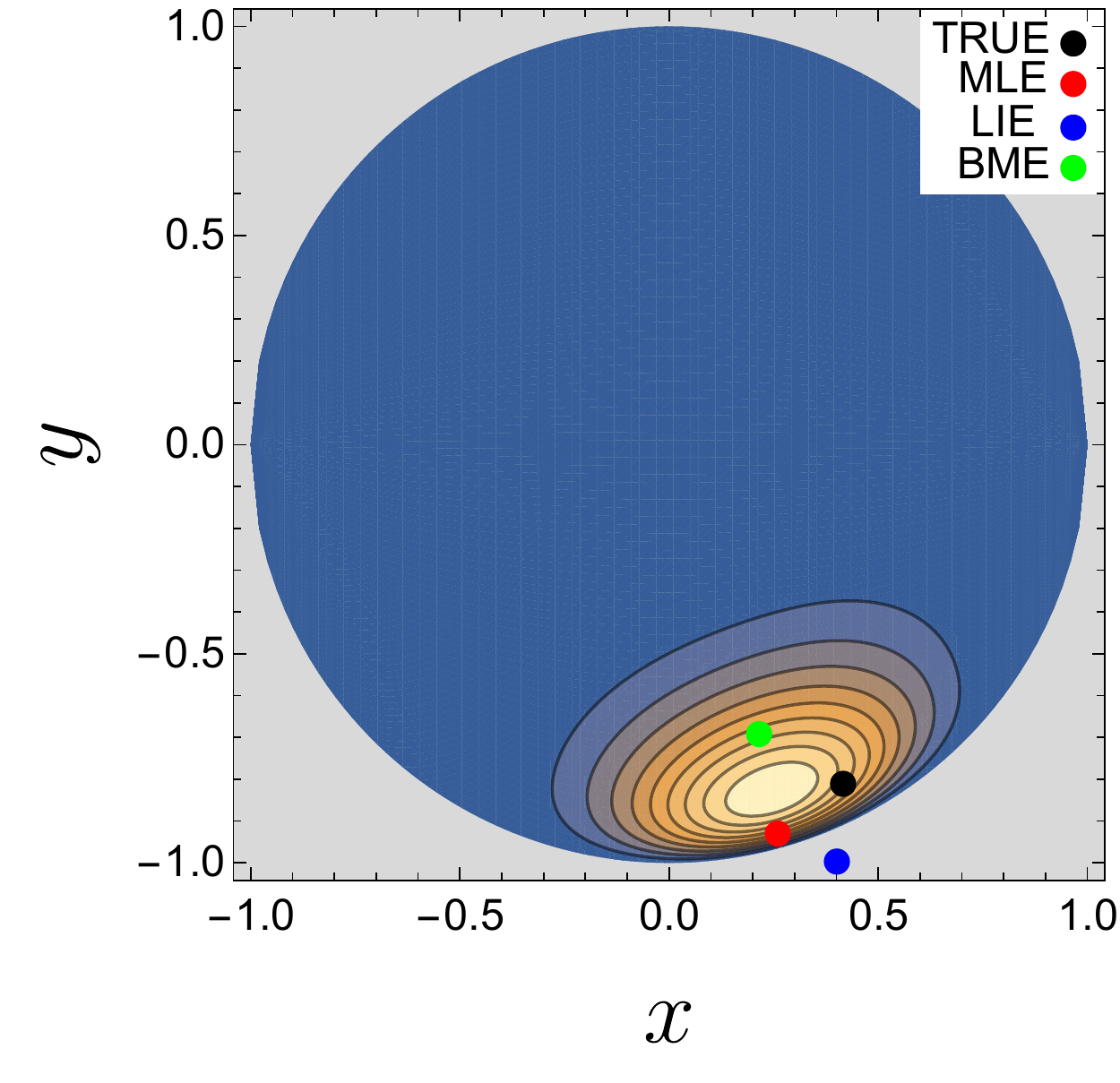}
\includegraphics[scale=0.26]{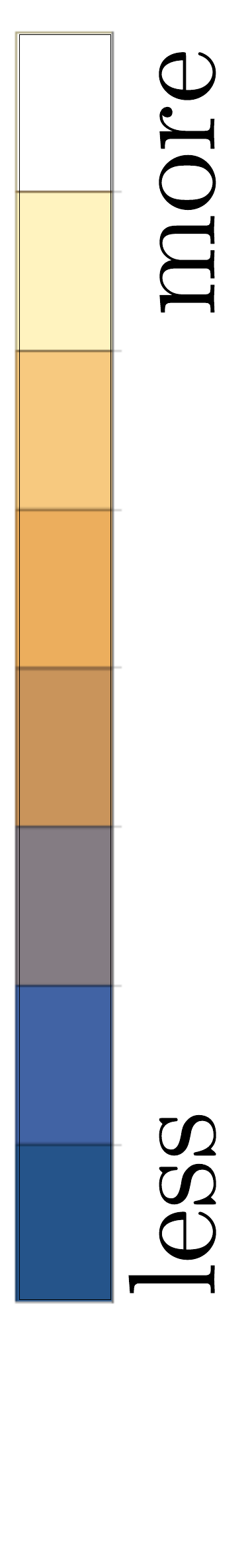}
\caption{We plot posterior marginal distributions $P(x,z|\mathcal{D})$, $P(y,z|\mathcal{D})$, and $P(x,y|\mathcal{D})$ at top left, top right, and bottom, respectively. The contour plots relate the relative probability of Bloch sphere coordinates. The plotted dots represent the locations of the true state, MLE, LIE, and BME. These plots are given to emphasize the physicality of the posterior distribution used to calculate the BME--the posterior distributions conform to quantum bounds. \label{bloch}}
\end{figure}

To illustrate the physicality of the distribution $P(\tau|D)$ and the differences between the MLE, LIE, and the BME consider a true quantum state $\rho_0$ defined by the parameters $u_0=0.864$, $\theta_0=0.393$, and $\phi_0=5.18$. For visualization we use the Bloch sphere where the state is represented by the expectations of the the Pauli operators given in Eq. \ref{pauliZ}-\ref{pauliY}. We simulated taking 10 measurements in the $Z$, $X$, and $Y$ bases from which we generated counts $c_h=7$, $c_v=3$, $c_d=7$, $c_a=3$, $c_l=0$, and $c_r=10$.  We plot the distributions $P(x,z|\mathcal{D})$, $P(y,z|\mathcal{D})$, and $P(x,y|\mathcal{D})$ in Fig. \ref{bloch}. The coordinates for each estimate are given in Table \ref{tableEstimates}. This small data set emphasizes the parametrized distribution's physicality and the qualitative difference between the MLE and BME. The gray locations correspond to unphysical states. As can be seen in the top right and bottom plots in Fig. \ref{bloch}, the LIE can be unphysical. To correct this, the MLE is found on the boundary, a pure state. In contrast, the BME will always be located within the physical space. This illustration is not made to emphasize the performance of any of specific approach. Performance is addressed in Section \ref{performance}.
\begin{table}[t]
\centering
\small
\begin{tabular}{|c|c|c|c|c|}
\hline
 & $z$ & $x$ & $y$ & $\sqrt{z^2+x^2+y^2}$ \\
\hline
true &-0.156&0.414&-0.813& 0.925\\\hline
MLE &0.263& 0.263& -0.928& 1.00\\\hline
LIE &0.400& 0.400& -1.00& 1.15$^\dagger$\\\hline
BME &0.226&0.216&-0.695&0.762\\\hline
\hline
\end{tabular}
\normalsize
\caption{Bloch sphere coordinates for the true state, MLE, LIE, and BME. $\dagger$In this case, the LIE is unphysical.\label{tableEstimates}}
\end{table}

In order to utilize the ideal single qubit formalism with single-qubit experiments, the data can be renormalized to the lowest efficiency measurement similar to the procedure used in Appendix \ref{mleAppendix}. This method does not fully utilize the available information and has the additional complication that preliminary experiments must transpire to determine the measurement efficiencies. In the remainder of our manuscript we address qubit estimation for experiments.

\section{Bayesian mean estimation for multi-qubit experiments}\label{Sec:twoqubits}
In an experiment the probability of observing an outcome depends not only on the quantum state but also on the measurement apparatus itself. In this case imperfections and asymmetries in the measurement process prohibit the type of ``perfect" estimate we investigated in the last section. Our experiment of investigation is the common two-photon experiment for which we introduce the fundamental assumptions and model below. James et al. \cite{measureQubits2001} previously reported an MLE approach to this experiment as well as higher dimensional experiments. In contrast to that method, we account for qubit loss within our defined likelihood and enable determination of the BME, the best estimate on average \cite{blume2010optimal}, which avoids MLE pitfalls such as ``zero" probabilities, impossible outcomes. 

\subsection{Estimating parameters in a single-basis two-photon experiment}\label{A}
\begin{figure}[tb]
\centering
\includegraphics[width=0.6\linewidth]{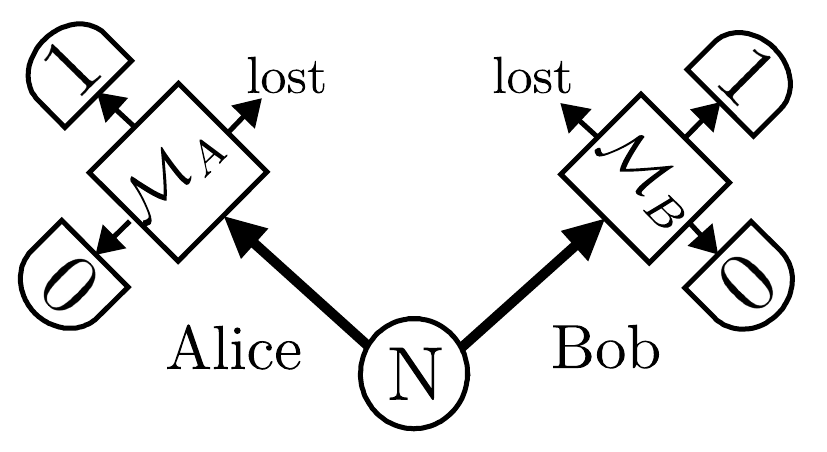}
\caption{The two-photon experiment is illustrated above. One member of a photon pair, a qubit, is sent to both Alice and Bob who have each chosen a measurement basis. An individual qubit can result in one of two orthogonal outcomes, $0$ or $1$, or the qubit can be lost.\label{setup}}
\end{figure}

\begin{figure}[tb]
\centering
\includegraphics[width=\linewidth]{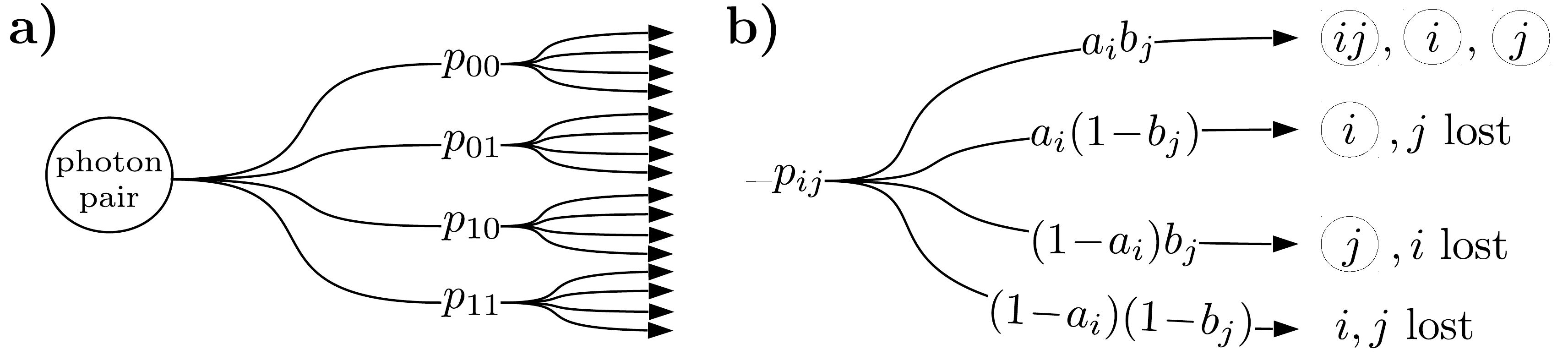}
\caption{a) Our Bayesian tree begins with the existence of a photon pair. This pair is then ``destined" to the joint outcome $ij$ according to probability $p_{ij}$. b) A closer view of each tree branch shows that each has four possible terminations due to  pathway inefficiencies. These possibilities include a joint event, a single event (one photon is lost), and no event (both photons are lost).\label{bayes_tree}}
\end{figure}

In this section, we give an example of estimating parameters in a single-basis experiment. To begin, we assume the existence of a photon pair. A member of this pair is sent to Alice and the other one to Bob each of whom has chosen a measurement basis as seen in Fig. \ref{setup}. A single photon can result in one of two observable orthogonal outcomes, $0$ or $1$, and one unobservable outcome, the photon is lost. All observable outcomes have probabilities of occurrence proportional to the joint probabilities $p_{00}$, $p_{01}$, $p_{10}$, and $p_{11}$ as seen in Fig. \ref{bayes_tree}a. Additionally, Fig. \ref{bayes_tree}b illustrates the four possible outcomes for a given ``destiny" when pathway efficiencies are considered. The possibilities include both photons being counted giving one coincidence count and two singles counts, one photon being counted and one lost giving one singles count, or both photons being lost giving no counts. 

Alice and Bob record event $0$ or event $1$ with number $A_0,A_1\leq N$ and $B_0,B_1\leq N
$, respectively, since typically some portion of the $N$ photons are lost. Losses are due to Alice and Bob's suboptimal pathway efficiencies $\left\{a_0,a_1,b_0,b_1\right\}\in\left[0,1\right]$. In the event both members of a photon pair are detected, Alice and Bob observe joint results, giving coincidence totals $c_{00},c_{01},c_{10},$ and $c_{11}$.

From this data we may enumerate the number of each type of event. The number of joint coincidence events are straightforward, given by the $c_{ij}$ with the probability of these events being $a_i b_j p_{ij}$. The number of events where Alice registers result $i$ and Bob loses his photon is $A_i\mst c_{i0}\mst c_{i1}$. The probability of this occurrence is $a_i\left[(1-b_0)p_{i0}+(1-b_1)p_{i1}\right]$. The terms for Bob registering a photon and Alice losing her photon are similar. The number of events where both photons are lost is $N\mst A_{0}\mst A_{1}\mst B_{0}\mst B_{1}\+ c_{00}\+ c_{01}\+ c_{10}\+ c_{11}$ with probability
\begin{equation}p_{\substack{pair\\lost}}=(1\mst a_0)(1\mst b_0)p_{00}+(1\mst a_0)(1\mst b_1)p_{01}+(1\mst a_1)(1\mst b_0)p_{10}+(1\mst a_1)(1\mst b_1)p_{11}\textrm{.}\nonumber\end{equation}

For now, assume the photon number $N$ is known. In this case, using Bayes' rule, the event number, and the probabilities given above, the PD is
\begin{equation}P\left(\alpha|\mathcal{D},N\right)=\frac{P\left(\mathcal{D},N|\alpha\right)P(\alpha)}{P\left(\mathcal{D},N\right)}\end{equation}
where $\alpha$=$\left\{p_{00},p_{01},p_{10},p_{11},a_0,a_1,b_0,b_1\right\}$ are the unknown parameters, joint probabilities and pathway efficiencies, the data set $\mathcal{D}$=$\left\{c_{00},c_{01},c_{10},c_{11},A_0,A_1,B_0,B_1\right\}$ consists of the known singles and coincidence count values totalling 
\begin{equation}s=A_0+A_1+B_0+B_1\quad\quad\quad n=c_{00}+c_{01}+c_{10}+c_{11}\textrm{,}\nonumber\end{equation}
respectively,
\begin{align} &P(\mathcal{D},N|\alpha)=\nonumber\\
&\gamma\!\left(N\right)(a_0b_0p_{00})^{c_{00}}(a_0b_1p_{01})^{c_{01}}(a_1b_0p_{10})^{c_{10}}(a_1b_1p_{11})^{c_{11}}\nonumber\\
&\qquad\times [a_0\left(p_{00}(1-b_0)+p_{01}(1-b_1)\right)]^{A0-c_{00}-c_{01}}[a_1\left(p_{10}(1-b_0)+p_{11}(1-b_1)\right)]^{A1-c_{10}-c_{11}}\nonumber\\
&\qquad\times [b_0\left(p_{00}(1-a_0)+p_{10}(1-a_1)\right)]^{B0-c_{00}-c_{10}}[b_1\left(p_{01}(1-a_0)+p_{11}(1-a_1)\right)]^{B1-c_{01}-c_{11}}\nonumber\\
&\qquad\times[p_{00}(1\mst a_0)(1\mst b_0)+p_{01}(1\mst a_0)(1\mst b_1)+p_{10}(1\mst a_1)(1\mst b_0)+p_{11}(1\mst a_1)(1\mst b_1)]^{N-(s-n)}\textrm{,}\nonumber\\
&P(\alpha)=1\textrm{,}\nonumber\\
&P(\mathcal{D},N)=\!\!\!\!\int\!\!d\alpha \;P(\mathcal{D},N|\alpha)P(\alpha)\textrm{,}\nonumber\\
&\gamma\!\left(N\right)=\frac{N!}{(N\!-\!(s\!-\!n))!(A_0\mst c_{00}\mst c_{01})!(A_1\mst c_{10}\mst c_{11})!(B_0\mst c_{00}\mst c_{10})!(B_1\mst c_{01}\mst c_{11})!c_{00}!c_{01}!c_{10}!c_{11}!}\textrm{.}\nonumber\end{align}
The likelihood $P(\mathcal{D},N|\alpha)$ consists of the probability of each type of event with a multiplicity equal to the number of times it occurred. Both the probabilities and number of events were described in the preceding paragraph. We have retained the full form of the likelihood that includes $\gamma(N)$ for use below.

It is typical in two-photon experiments that the photon number $N$ is not known. If $N$ is known, the following step may be skipped and the above PD is the appropriate choice. Otherwise, we must make $N$ an unobserved parameter or seek a way to eliminate it. Fortunately, there is an analytical method to remove $N$ from the PD completely \cite{jaynes2003probability} by taking an average over the $N$ distribution using the summation formula
\begin{equation}\sum_{m=0}^\infty\binom{m+y}{m}m^zx^m=\left(x\frac{d}{dx}\right)^z(1-x)^{-(y+1)}\textrm{.}\label{sumOverN}\end{equation}
Since the average is taken over the distribution, see Fig. \ref{estimates}, only probable values of $N$ will have appreciable contribution. Applying this formula, $N$ is removed giving PD
\begin{equation}P\left(\alpha|\mathcal{D}\right)=\frac{\sum_{N=s-n}^{\infty}P\left(\mathcal{D},N|\alpha\right)P(\alpha)}{P\left(\mathcal{D}\right)}=\frac{P\left(\mathcal{D}|\alpha\right)P(\alpha)}{P\left(\mathcal{D}\right)}\label{PD}\end{equation}
where
\begin{align}& P(\mathcal{D}|\alpha)=a_0^{A_0}a_1^{A_1}b_0^{B_0}b_1^{B_0}p_{00}^{c_{00}}p_{01}^{c_{01}}p_{10}^{c_{10}}p_{11}^{c_{11}}\nonumber\\
&\qquad\times[p_{00}(1-b_0)+p_{01}(1-b_1)]^{A0-c_{00}-c_{01}}[p_{10}(1-b_0)+p_{11}(1-b_1)]^{A1-c_{10}-c_{11}}\nonumber\\
&\qquad\times[p_{00}(1-a_0)+p_{10}(1-a_1)]^{B0-c_{00}-c_{10}}[p_{01}(1-a_0)+p_{11}(1-a_1)]^{B1-c_{01}-c_{11}}\nonumber\\
&\qquad\times[1\mst p_{00}(1\mst a_0)(1\mst b_0)\mst p_{01}(1\mst a_0)(1\mst b_1)\mst p_{10}(1\mst a_1)(1\mst b_0)\mst p_{11}(1\mst a_1)(1\mst b_1)]^{-s+n-1}\textrm{,}\label{likelihood2}\\
&P(\alpha)=1\textrm{,}\nonumber\\
&P(\mathcal{D})=\int\!\!d\alpha\;P(\mathcal{D}|\alpha)P(\alpha)\textrm{,}\nonumber\end{align}
and, in this specific case, 
\begin{equation}\int\! d\alpha \!\equiv\! \!\int_0^1\!\!\!\!da_0\!\int_0^1\!\!\!\!da_1\!\int_0^1\!\!\!\!db_0\!\int_0^1\!\!\!\!db_1\!\!\int_{0}^1\!\!\!\! dp_{00}\!\! \int_{0}^{1-p_{00}}\hspace{-25pt}dp_{01}\!\!\int_{0}^{1-p_{00}-p_{01}}\hspace{-42pt}dp_{10}\nonumber \end{equation}
with $p_{11}=1-p_{00}-p_{01}-p_{10}$. We omitted all constants.

Assuming the integral can be carried out, we can make estimates of any parameter via its the mean value, for instance,
\begin{equation}\overline{p_{00}}=\int\!\!d\alpha P\left(\alpha|\mathcal{D}\right)\times p_{00}.\end{equation}
Likewise, any other parameter mean $\;\overline{p_{ij}}\;$, $\;\overline{a_{i}}\;$, or $\;\overline{b_{i}}\;$ as well as their standard deviations may be estimated. One exception is the mean value $\overline{N}$. We find this mean by setting $z=1$ in Eq. (\ref{sumOverN}),
\begin{equation}\overline{N}=\int\!\!d\alpha P\left(\alpha|\mathcal{D}\right)\times\frac{s-n+g(\alpha)}{1-g(\alpha)}\end{equation}
where 
\begin{equation}g(\alpha)=p_{00}(1\mst a_0)(1\mst b_0)+p_{01}(1\mst a_0)(1\mst b_1)+p_{10}(1\mst a_1)(1\mst b_0)+p_{11}(1\mst a_1)(1\mst b_1)\textrm{.}\end{equation}

In principal, all of the above integrals have analytical solutions via the multinomial theorem,
 \begin{equation}(x_0+x_1+...+x_m)^n =\sum_{k_0+k_1+...+k_m=n}\!\!\binom{n}{k_0,k_1,..,k_n}x_0^{k_0}x_1^{k_1}\cdots x_m^{k_m}\textrm{,}\nonumber\end{equation}
which gives exact answers in the form of sums of Beta and Gamma functions. However, the computation needed to carry out the resultant sums is prohibitive. 

If we cannot efficiently make our parameter estimations analytically, we can utilize numerical sampling to approximate the BMEs of interest. We discuss this in detail in Appendix \ref{ss}. If the probability of obtaining a sample $\alpha^{(r)}$ tends to the true probability $P(\alpha^{(r)}|\mathcal{D})$, the mean estimations can then be made by repetitive sampling, 
\begin{align}\overline{\alpha}=\frac{1}{R}\sum_{r=0}^R\alpha^{(r)}&=\frac{1}{R}\sum_{r=0}^R\left\{p_{00}^{(r)},p_{01}^{(r)},p_{10}^{(r)},p_{11}^{(r)},a_0^{(r)},a_1^{(r)},b_0^{(r)},b_1^{(r)}\right\}\nonumber\\
&=\left\{\overline{p_{00}},\;\overline{p_{01}},\;\overline{p_{10}},\;\overline{p_{11}},\;\overline{a_{0}},\;\overline{a_{1}},\;\overline{b_{0}},\;\overline{b_{1}}\right\}\textrm{.}\nonumber\end{align}

\subsection{Single-basis simulation}
\begin{figure}[b]
\centering
\includegraphics[width=0.32\linewidth]{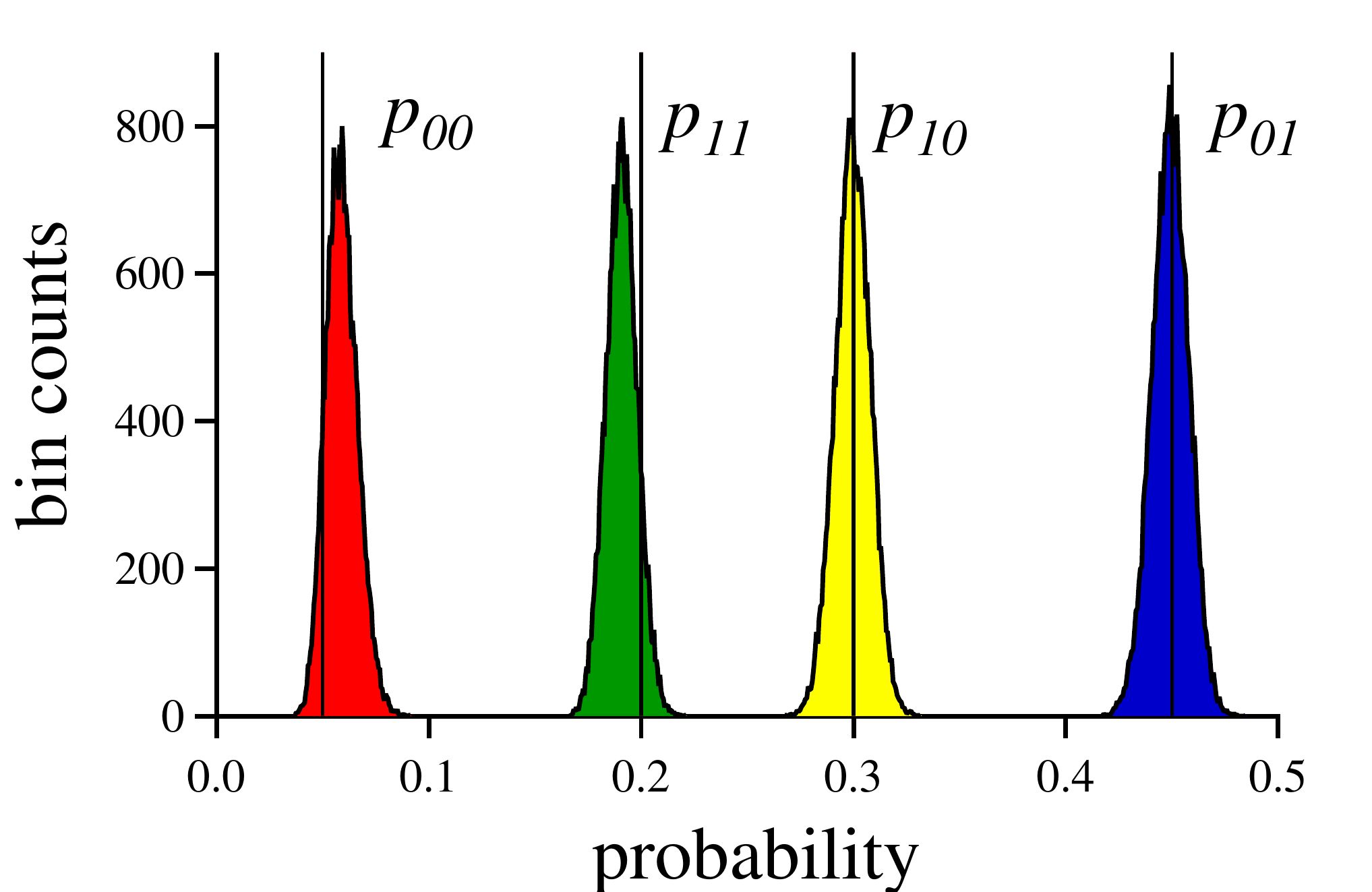}
\includegraphics[width=0.32\linewidth]{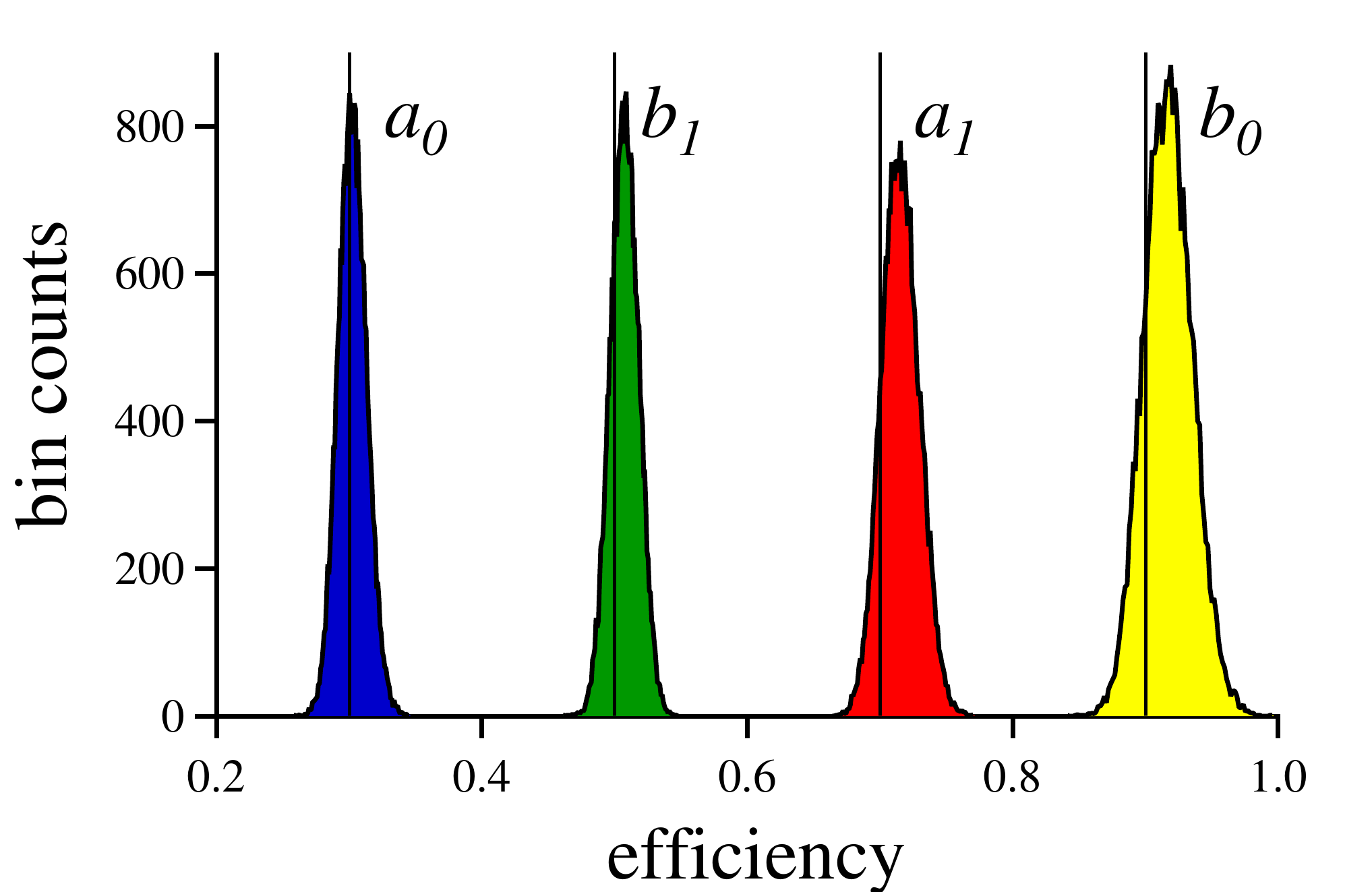}
\includegraphics[width=0.32\linewidth]{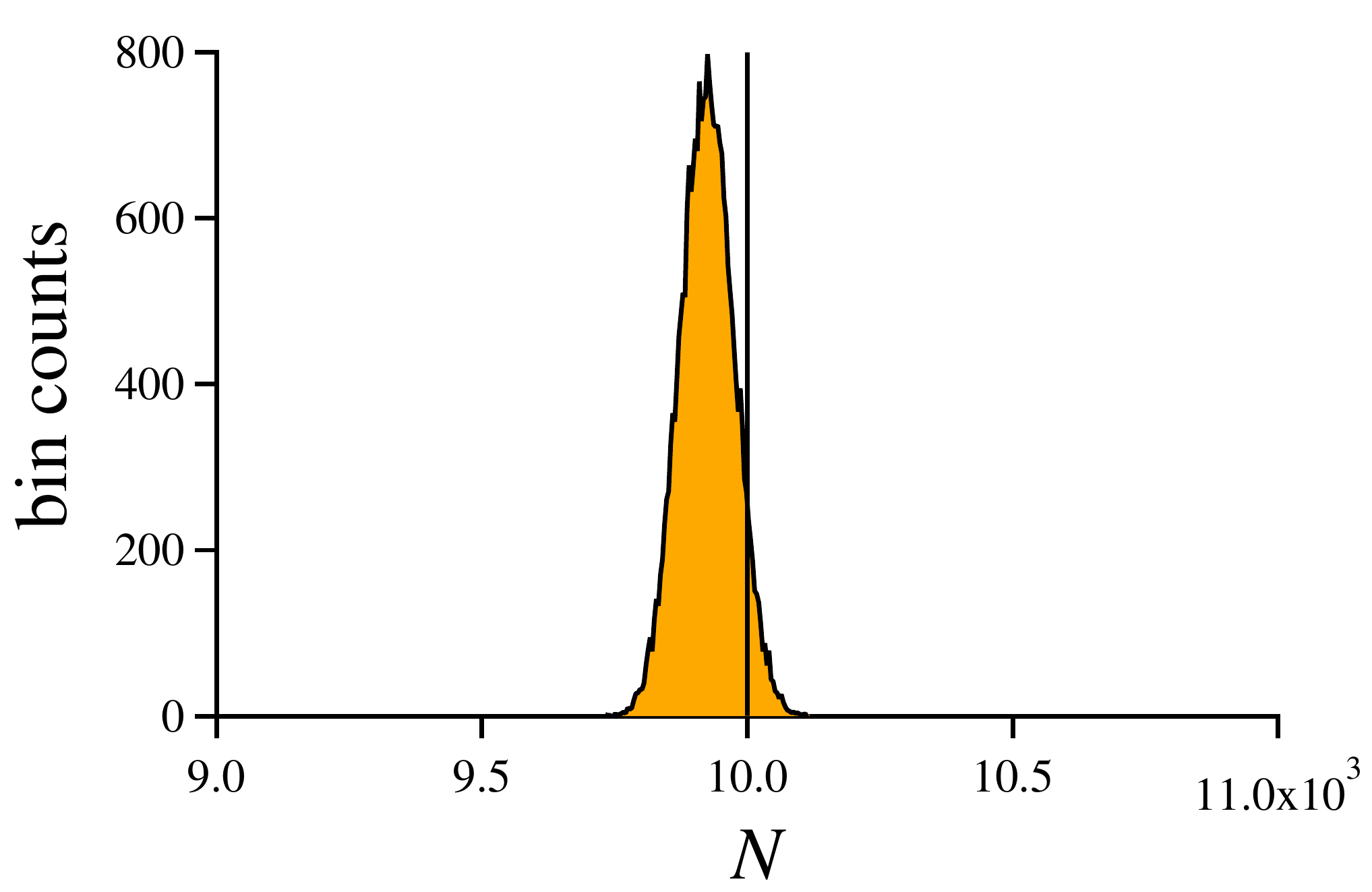}
\caption{Left. Sample histograms given proportional to the approximate the probability distribution for each parameter $p_{ij}$ whose true value is given by the black vertical line. Middle. Sample histograms are given proportional to the probability distribution for each efficiency parameter $a_{0}$, $a_{1}$, $b_{0}$, and $b_{1}$ whose true value is given by the black vertical line. Right. A histogram of samples for parameter $N$ are given proportional to the probability distribution for photon number $N$.\label{estimates}}
\end{figure}
Consider a single-basis simulation where unbeknownst to Alice and Bob a source generates $N$=$10,000$ photon pairs with joint probabilities and pathways efficiencies
\begin{align}p_{00}&=0.3\quad\quad p_{01}=0.05\quad\quad p_{10}=0.2\quad\quad p_{11}=0.45\nonumber\\
a_{0}&=0.3\quad\quad\; a_{1}=0.7\quad\quad \;\;\;b_{0}=0.9\quad\quad \;\;b_{1}=0.5\;\textrm{.}\nonumber\end{align}
The only information available to Alice and Bob are their count numbers
\begin{align}A_0&=1079\quad\quad A_{1}=4553\quad\quad B_{0}=4474\quad\quad B_{1}=2565\nonumber\\
c_{00}&=829\phantom{0}\quad\quad c_{01}=89\phantom{00}\quad\quad c_{10}=1245\quad\quad c_{11}\!=1624\nonumber\textrm{.}\end{align}
Numerical sampling, see Appendix \ref{ss}, is used to produce sample $\alpha^{(r)}$ from $P\left(\alpha|\mathcal{D}\right)$. Fig. \ref{estimates} includes histograms for each parameter from 25,200 $\alpha$ samples. Each parameter histogram contains 100 bins. This large sample size was chosen to illustrate that the samples do come from a distribution. For the typical application a much smaller sample size would likely be adequate.  From the distribution $P\left(\alpha|\mathcal{D}\right)$ parameter mean values are found to be
\begin{align}\overline{p_{00}}=0.300\pm 0.008& \quad\overline{p_{01}}=0.059 \pm 0.007 \quad \overline{p_{10}}=0.191\pm0.007&&\quad\overline{p_{11}}=0.450\pm0.009\nonumber\\
\overline{a_{0}}=0.303\pm0.011&\quad \overline{a_{1}}=0.716\pm0.014 \quad \overline{b_{0}}=0.918\pm0.018&&\quad \overline{b_{1}}=0.508\pm0.011\nonumber\end{align}
and mean photon number $\overline{N}=9926\pm 50.9$. Comparison with the above true values shows qualitative agreement.

\subsection{Parametrizing the n-dimensional density matrix}\label{C}
For approachability, we obscured the construction of the density matrix for the ideal single qubit given in Eq. \ref{ideal_rho}. We briefly describe this construction here, for a full proof with discussion see \cite{daboul1967conditions}. We note that this construction is similar to that recently proposed by Seah et al. \cite{MCsamplingQStates_II} whose density matrix sampling application is similar to our approach. Daboul's parametrization can be extended to quantum systems of any dimension. For an $n$-dimensional Hilbert space, the density matrix is formed using the Cholesky decomposition, requiring $\rho=L(\!\tau\!)L(\!\tau\!)^\dagger$ with
\begin{equation}L(\!\tau\!)\! =\!\!\left(\!\begin{array}{ccccc}
    L_{11} (\!\tau\!)\!\!& 0& 0 & \cdots& 0 \\
    L_{21}(\!\tau\!)\! \! & L_{22}(\!\tau\!)\!\! & 0 & \cdots& 0 \\
    L_{31}(\!\tau\!)\! \! & L_{32}(\!\tau\!)\!\!& L_{33}(\!\tau\!)\!\! & \cdots& 0 \\
    \vdots&     \vdots&     \vdots & \ddots& \vdots \\
    L_{n1}(\!\tau\!)\!\! & L_{n2}(\!\tau\!)\!\! & L_{n3}(\!\tau\!)\!\! &\cdots& \!\!\! L_{nn}(\!\tau\!) \end{array}\right)\nonumber\vspace{5pt}\end{equation}
being a lower triangular matrix with positive real diagonal elements. The parameter set $\tau$ include $n^2$$-$$1$ parameters which describe a unique density matrix.
 
The elements $L_{ij}$ may be written as
\begin{align} L_{ij} (\!\tau\!)&=U_i V_{ij}\phantom{0}\quad\quad (j\leq i)\nonumber \\
L_{ij} (\!\tau\!)&=0\phantom{U_i V_{ij}}\quad\quad (j> i)  \nonumber \end{align}
where

\begin{align}U_{1}&=\cos\left(u_1\right) \hspace{0.3\linewidth} V_{ii}=1 \nonumber \\
U_{k}&=\cos\left(u_k\right)\prod_{j=1}^{k-1}\sin\left(u_j\right)\quad \!\!\textrm{\scriptsize $(1<k<n)$} \;\hspace{0.05\linewidth} V_{i1}=\cos\left(\theta_{i1}\right)e^{i\phi_{i1}}\quad (i>1)\nonumber \\
U_{n}&=\prod_{j=1}^{n-1}\sin\left(u_j\right) \hspace{0.26\linewidth} V_{ik}=\cos\left(\theta_{ik}\right)e^{i\phi_{ik}}\prod_{j=1}^{k-1}\sin\left(\theta_{ij}\right)\quad \!\!\textrm{\scriptsize $(1<k<i)$} \textrm{.}\nonumber  \end{align}

Consider the case of two qubits with dimension $n$=$4$, the parametrized matrix elements of $L(\!\tau\!)$ are
\begin{align}L_{11}(\!\tau\!)\!&=\!\cos(u_1)\nonumber\\
    L_{21}(\!\tau\!)\!&=\!\sin(u_1)\cos(u_2)\cos(\theta_{21})e^{i \phi_{21}}\nonumber\\
    L_{22}(\!\tau\!)\!&=\!\sin(u_1)\cos(u_2) \sin(\theta_{21})\nonumber\\
        L_{31}(\!\tau\!)\!&=\!\sin(u_1)\sin(u_2)\cos(u_3)\cos(\theta_{31})e^{i \phi_{31}}\nonumber\\
L_{32}(\!\tau\!)\!&=\!\sin(u_1)\sin(u_2)\cos(u_3)\sin(\theta_{31})\cos(\theta_{32})e^{i \phi_{32}}\nonumber\\
L_{33}(\!\tau\!)\!&=\!\sin(u_1)\sin(u_2)\cos(u_3)\sin(\theta_{31})\sin(\theta_{32})\nonumber\\
        L_{41}(\!\tau\!)\!&=\!\sin(u_1)\sin(u_2)\sin(u_3)\cos(\theta_{41})e^{i \phi_{41}}\nonumber\\
        L_{42}(\!\tau\!)\!&=\!\sin(u_1)\sin(u_2)\sin(u_3)\sin(\theta_{41})\cos(\theta_{42})e^{i \phi_{42}}\nonumber\\
        L_{43}(\!\tau\!)\!&=\sin(u_1)\sin(u_2)\sin(u_3)\sin(\theta_{41})\sin(\theta_{42})\cos(\theta_{43})e^{i \phi_{43}}\nonumber\\
        L_{44}(\!\tau\!)\!&=\!\sin(u_1)\sin(u_2)\sin(u_3)\sin(\theta_{41})\sin(\theta_{42})\sin(\theta_{43})
\nonumber
\end{align}
with $u_{i}\in[0,\frac{\pi}{2}]$, $\theta_{ij}\in[0,\frac{\pi}{2}]$, and $\phi_{ij}\in[0,2\pi]$. Indeed, one could instead change the $u_i$ and $\theta_{ij}$ trigonometric terms to 
\begin{equation}\cos(u_i)\rightarrow \sqrt{u_i'}\quad\quad
\sin(u_i)\rightarrow \sqrt{1-u_i'}\quad\quad
\cos(\theta_i)\rightarrow \sqrt{\theta_i'}\quad\quad
\sin(\theta_i)\rightarrow \sqrt{1-\theta_i'}\nonumber 
\end{equation}
with $u_i'\in[0,1]$, $\theta_{ij}'\in[0,1]$. The complex terms involving $\phi_{ij}$ remain unchanged. A similar adjustment was used by Chung and Trueman \cite{ChungTrueman}. 

\subsection{Estimating parameters in a multi-basis two-photon experiment}\label{multi}
In section \ref{A} and \ref{C}, respectively, we defined our experimental likelihood for the single-basis experiment and detailed the parametrization of any $n$-dimensional density matrix. To make estimations using data from multi-basis two-photon experiment we will use both of these pieces. Our example will be full-state tomography. Other multiple basis experiments will have similar estimation constructions. In the case that the data set is incomplete, our method will still return an estimate true to both the given data and all quantum constraints.

To complete full-state tomography Alice and Bob each take measurements in bases $Z$, $X$, and $Y$ such that all outcomes are observable in each basis combination $ZZ$, $ZX$, $XZ$, $ZY$, $YZ$, $XX$, $XY$, $YX$, and $YY$. Thus, Alice and Bob's data set will include the data from all 9 basis combinations. The likelihood is a product of the single-basis likelihoods, Eq. \ref{likelihood2}, from each of these basis combinations,
\begin{equation}P(\mathcal{D}|\alpha)\!=\!P(\mathcal{D}_{ZZ}|\alpha_{ZZ})P(\mathcal{D}_{ZX}|\alpha_{ZX})\cdots P(\mathcal{D}_{YY}|\alpha_{YY})\label{likelihood_product}\end{equation}
where $\alpha$ includes the probabilities of all measurement outcomes and the four experimental pathway efficiencies which we assume are the same over all bases. However, in the experimental section, Section \ref{Sec:Experimental}, we do not make this assumtion. Next, we parametrize our density matrix using the hyperspherical parameters described in the previous section,
\begin{equation}P(\mathcal{D}|\alpha)\rightarrow P(\mathcal{D}|\tau)\textrm{.}\end{equation}
This parametrization comes with a new measure defined by Eq. \ref{measure}, \ref{dTau}, and \ref{gIJ}.

Putting it all together we can make any BME of interest, for instance the mean density matrix
\begin{equation}\overline{\rho}\;= \frac{1}{P(\mathcal{D})}\int d\tau P(\mathcal{D}|\tau) P(\tau) \times \rho(\tau) \label{hardintegral}\end{equation}
where $P(\mathcal{D})=\int d\tau  P(\mathcal{D}|\tau) P(\tau)$.

If it is not computationally convenient to evaluate the integrals of the type given in Eq. \ref{hardintegral}, we can utilize numerical sampling. If we can draw samples $\rho^{(r)}$ from the distribution $P(\tau|\mathcal{D})$ we can estimate the BME of our density matrix as 
\begin{equation}\overline{\rho}\;= \lim_{R\rightarrow\infty}\frac{1}{R} \sum_{i=1}^{R} \rho^{(r)}\label{PDlast}\textrm{.}\end{equation}
We address our numerical sampling approach in Appendix \ref{ss}.

\subsection{State certainty}
When reporting the values of experimental measurements such as the visibility of an interference curve $V$ or the value of the Bell parameter $S$, it is typical to provide a standard deviation to describe the uncertainty in the parameter, e.g. $V=0.98\pm 0.01$ or $S=2.65\pm 0.05$. This gives a quantification of the uncertainty in the estimate. When the BME is multi-dimensional the uncertainty can be represented by a covariance matrix \cite{blume2010optimal,Granade2016}
\begin{equation}\Delta \rho(\tau)\!=\!\left(\!\begin{array}{cccc}
    \Delta \tau_0^2 & \Delta \tau_0 \tau_1 & \cdots &\Delta \tau_0 \tau_k\\
    \Delta \tau_1 \tau_0 & \Delta \tau_1^2 & \cdots &\Delta \tau_1 \tau_k\\
    \vdots & \vdots & \ddots & \vdots\\
    \Delta \tau_2 \tau_0 & \Delta \tau_2 \tau_1 & \cdots &\Delta \tau_k^2\\
    \end{array}\!\right)\end{equation}
with each element being a covariance,
\begin{equation}\Delta \tau_i \tau_j= \overline{\tau_i \tau_j}-(\overline{\tau_i})(\overline{\tau_j})\end{equation}
where $\overline{\tau_i}$ is the expectation, mean value, of $\tau_i$. For $i=j$ this is just the usual variance. Here the $k$ parameters include the $n^2-1$ parameters needed to define the density matrix as well as any additional experimental parameters such as the efficiencies.

We can also define a single quantity that captures a compact representation of the certainty in the estimation. We use the \emph{trace distance deviation} $\Delta D$, which is the mean trace distance over the distribution with the distribution's mean density matrix $\overline{\rho}$,
\begin{equation}\Delta D=\int d\tau D\textrm{\large$($}\;\overline{\rho},\rho(\tau)\textrm{\large$)$}\textrm{.}\label{dd}\end{equation}
The trace distance is
\begin{equation}D(\rho,\sigma)=\textrm{Tr}\left(\sqrt{\left(\rho-\sigma\right)^2}\right)=\frac{1}{2}\sum_{i}\left|\lambda_i\right|\label{traceD}\end{equation}
where the $\lambda_i$ are the eigenvalues of $\rho-\sigma$. We approximate $\Delta D$ with numerical sampling using the formula
\begin{equation}\Delta D\approx\frac{1}{R}\sum_{r=1}^R D\textrm{\large$($}\;\overline{\rho},\rho^{(r)}\textrm{\large$)$}\textrm{.}\end{equation}
When the certainty is high all samples will be close to the mean value giving $\Delta D\rightarrow 0$ which can be compared to the typical standard deviation where smaller is better. This is also useful when there is no particular state one wishes to compare the estimations with such as is typically done when reporting the fidelity. 

\section{BME performance with numerical sampling}\label{performance}
\begin{figure}[tbh]
\centering
\includegraphics[width=0.6\linewidth]{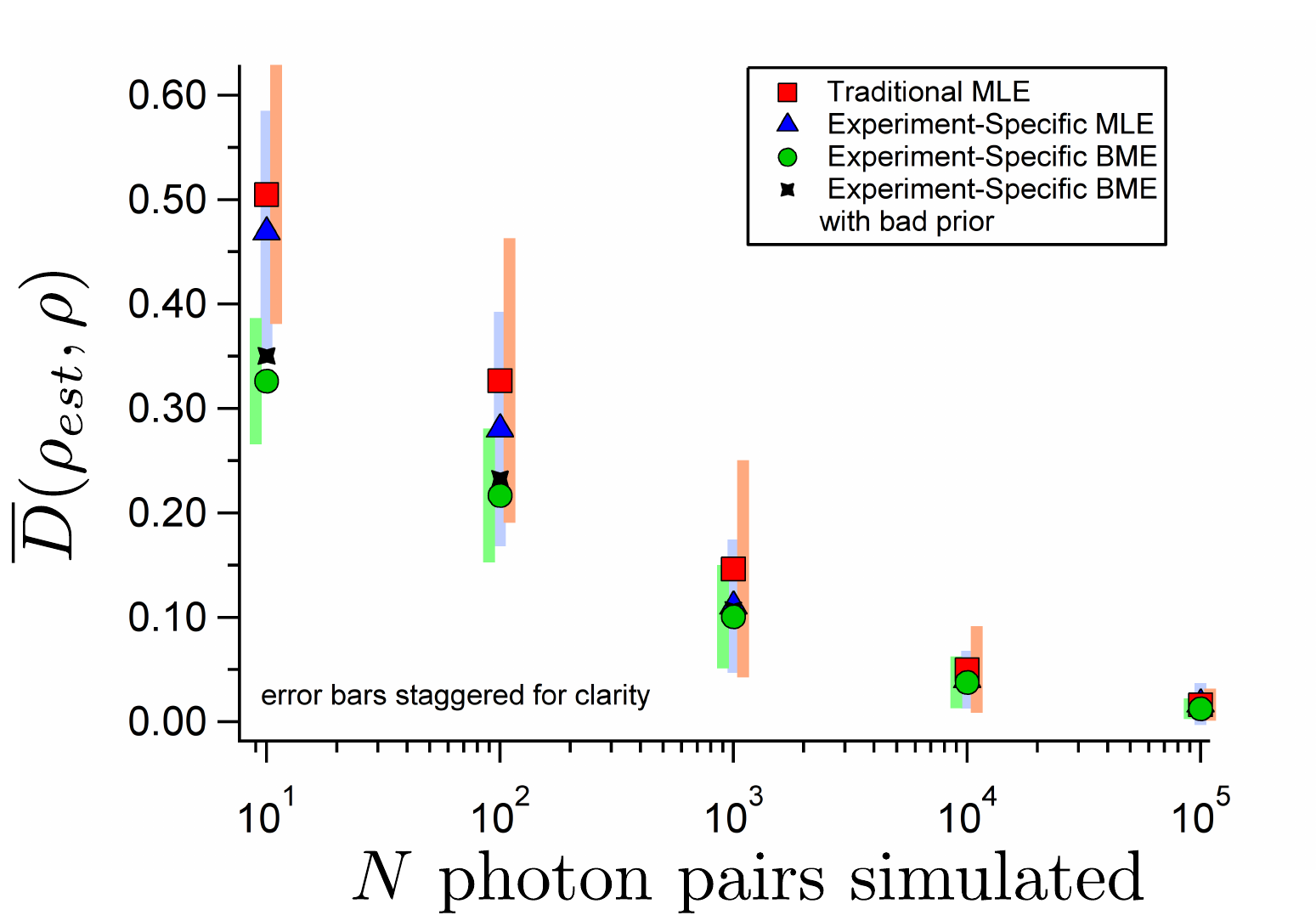}
\includegraphics[width=0.6\linewidth]{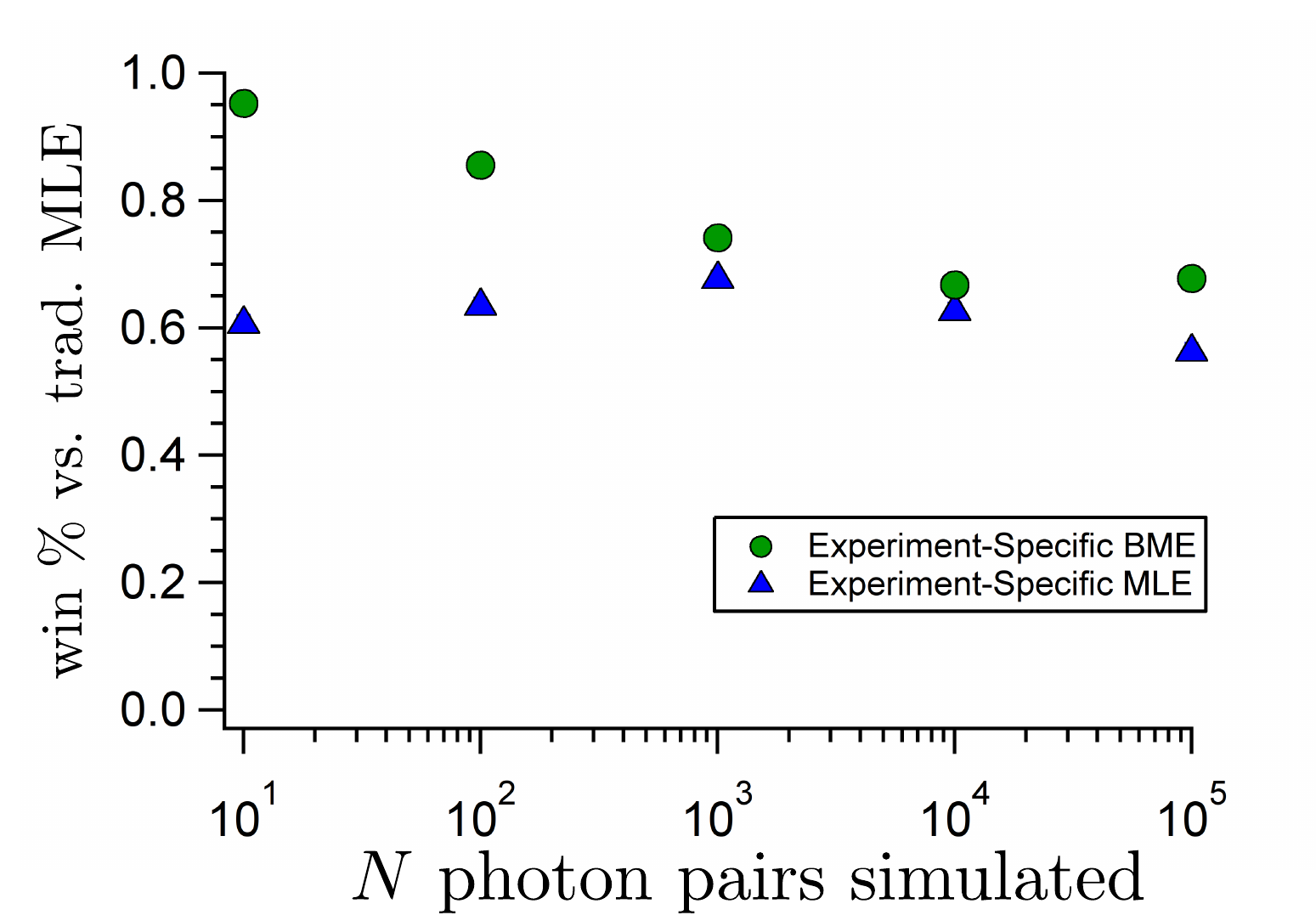}
\caption{We generated data from simulating two-photon experiments for various states and photon pair number $N$ as outlined in this section. Top. We have plotted the average distance estimate for photon pair number $N$ over 1000 randomly sampled states. Bottom. We give the win percentage for the experiment-specific BME and MLE versus the traditional maximum likelihood method. The best performance is achieved using the experiment-specific Bayesian mean estimate. Another observation from this data is that an experimentalist can achieve a better estimate by switching to an experiment-specific likelihood which allows them to forgo any preliminary experiments to determine normalizing constants.\label{results}}\end{figure}

To characterize the performance of the presented estimation methods we used the following procedure. For each $N\in\{10,10^2,10^3,10^4,10^5\}$ the following steps are repeated:
\begin{enumerate}[1.]
\item A density matrix $\rho$ is sampled from a uniform distribution using a Haar measure in the hypershperical parameter space.
\item A random set of pathway efficiencies $a_0, a_1, b_0,$ and $b_1$ are chosen from the range $\left[0,1\right]$. These are chosen to be the same across all bases.
\item We simulate a two-photon experiment for $N$ identical states $\rho$ in each of the 9 bases given in Section \ref{multi}, $9N$ total identical states to generate a data set $\mathcal{D}$. Each simulated experiment for a single basis is the same as described by the Bayesian tree in Section \ref{A}.
\item Using $\mathcal{D}$ we find  using a traditional likelihood and the actual randomly chosen pathway efficiencies as described in Appendix~\ref{mleAppendix}. Also with $\mathcal{D}$, we find the BME and MLE using the experiment-specific likelihood in which the pathway efficiencies are not known as described in Section \ref{multi}. Thus, the traditional MLE has the unfair and unrealistic advantage of knowing the pathway efficiencies exactly.
\item The distance $D$, Eq. \ref{traceD}, is found between each estimate and the true state $\rho$.
\item If the experiment-specific BME or MLE is closer to the true state than the traditional MLE, that estimation type has a win tallied.
\item Steps 1.-6. are repeated for $1000$ repetitions.
\item The average distance $\overline{D}$ over all $1000$ repetitions is found for the traditional MLE approach and the experiment-specific BME and MLE. The total wins versus the traditional MLE are also recorded.
\end{enumerate}
For these simulations the average distance $\overline{D}$ results are given at left in Fig.~\ref{results}, and the win percentages for the experiment-specific likelihood MLE and BME versus the traditional MLE are given at right in Fig.~\ref{results}. The MLE was found using an gradient ascent method described in Appendix~\ref{searchAppendix}. We emphasize that these results are conservative, since we give the traditional MLE process the pathway efficiencies exactly--these would not be known exactly in an experiment.

The states $\rho$ were drawn from a uniform distribution which is Haar invariant when using the measure obtained from Eq. \ref{measure}, \ref{dTau}, and \ref{gIJ}. We also use the same distribution as a prior to compute our BME estimate. We also made estimates using a non-Haar invariant measure to evaluate the significance of prior selection. This results in a drastically different prior relative to that used in generating the random state. In Fig. \ref{results} the experiment-specific BME with the original advantageous prior and the ``bad" prior are both plotted. As can be seen, there is, possibly, a small gain using the advantageous prior for the smaller photon pair number estimations. But, it also highlights that the prior choice can be made effectively inconsequential given enough data \cite{jaynes2003probability}.
The prior certainly can improve the estimate when little data is available. Granade and colleagues discuss this in depth \cite{Granade2016}. 
\section{Experimental tomography}\label{Sec:Experimental}
We performed state tomography on a two-photon polarization entangled target state 
\begin{equation}\left|\Psi^+\right\rangle=\frac{1}{\sqrt{2}}\left(\left|H_A\right\rangle\otimes\left|V_B\right\rangle+\left|V_A\right\rangle\otimes\left|H_B\right\rangle\right)\end{equation}
generated by pumping a periodically poled potassium titanyl phosphate (PPKTP) nonlinear crystal inside a Sagnac loop with two counterpropagating 405nm pump beams \cite{SagnacSource,tamperSeal}. The two possibilities of Type II \footnote{The signal and idler photons are produced with orthogonal polarizations in Type II SPDC.}  spontaneous parametric downconverison (SPDC), either the clockwise or counter-clockwise beam generated a 810nm photon pair,  leads to a polarization entangled state output into the idler and signal modes received by Alice and Bob, respectively. Alice and Bob each choose a basis by inclusion or omission of waveplates. Since this requires a physical adjustment to our apparatus for each basis choice, we assume in our likelihood that the efficiencies are independent parameters in each basis. The half-wave and quarter-wave plate matrix operations are, respectively,
\begin{equation}\textrm{H}=\left(
\begin{array}{cc}
    \frac{1}{\sqrt{2}} & \frac{1}{\sqrt{2}} \\
    \frac{1}{\sqrt{2}} & \frac{-1}{\sqrt{2}}
    \end{array}\right) \quad
    \textrm{Q}=\left(
\begin{array}{cc}
    1 & 0 \\
    0 & i
    \end{array}\right)\textrm{.}\nonumber
\end{equation}
To measure in basis $Z$ Alice omits her waveplates. To measure in $X$ she includes the half-wave plate, she operates on her single-photon with $H$. Finally, to measure in the $Y$ basis she includes both waveplates, operating with $Q$ then $H$. Single-photon detectors record the detection mode, orthogonal outcomes 0 or 1 in each basis.
\begin{figure}[t]
\centering
\includegraphics[width=0.38\linewidth]{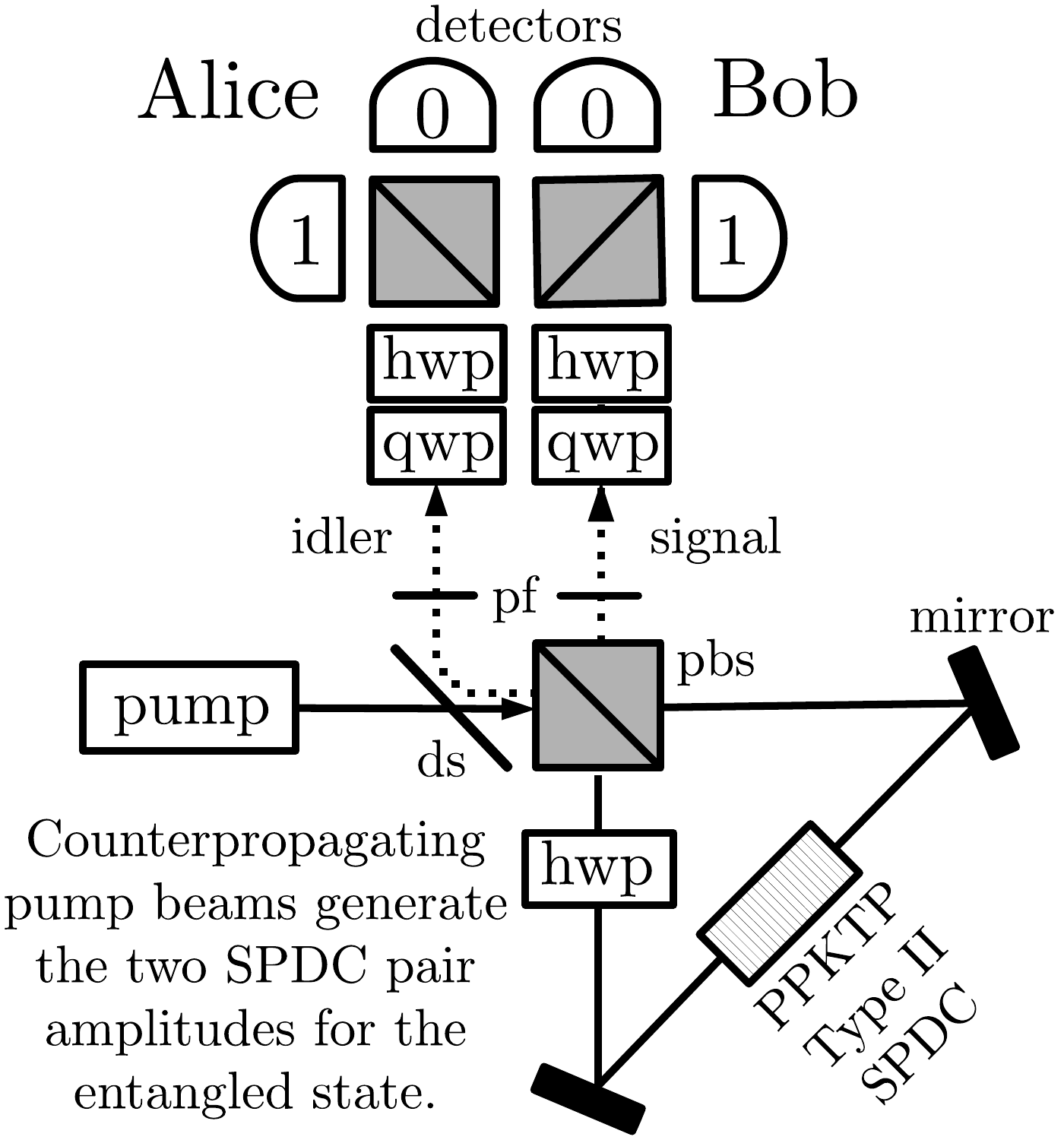}
\caption{Our two-photon polarization entangled state is generated by pumping a nonlinear PPKTP crystal inside a Sagnac loop with two counterpropagating pump beams each of which may generate Type II SPDC pairs. This leads to a polarization entangled state shared by Alice and Bob. Alice and Bob each choose a basis by inclusion or ommission of waveplates. Single-photon detectors record the detection mode, 0 or 1. ds$\equiv$dichroic splitter, pbs$\equiv$polarizing beamsplitter, pf$\equiv$pump filter, hwp$\equiv$ half-wave plate, qwp$\equiv$ quarter-wave plate\label{experiment}}
\end{figure}
From the experimental data given in Table 1, our mean density matrix is found to be
\begin{equation}\overline{\rho}\!=\!\!
\left(\!\begin{array}{cccc}
    0.01 & 0.03\+ i0.00 & 0.03\+ i0.00 & \mst 0.00\mst i0.01 \\
    0.02\mst i0.00 & 0.48 & 0.48\mst i0.02 &\mst 0.01\mst i0.04 \\
    0.03\mst i0.00 & 0.48\+ i0.02 & 0.49 & \mst 0.01\mst i0.05 \\
\textrm{-}0.00\+ i0.01 & \textrm{-}0.01\+ i0.04 & \mst 0.01\+ i0.05 & 0.02 \end{array}\!\right)\nonumber\end{equation}
with trace distance deviation $\Delta D=0.006$ defined in Eq. \ref{dd}. We have reported only 2 significant digits in $\;\overline{\rho}\;$ for brevity. Every element has a finite value, i.e. every outcome has a non-zero probability of occurrence. The fidelity of our mean $\overline{\rho}$ with the intended state $\Psi^+$ is
\begin{equation}\mathcal{F}=\sqrt{\left\langle \Psi^+ \right|\;\overline{\rho}\;\left|\Psi^+\right\rangle }=0.9838\pm0.0005 \textrm{.}\nonumber\end{equation}
We have not removed accidental coincidences from our estimation, we have assumed this contribution is negligible.

\begin{table}[t]
\centering
\small
\begin{tabular}{|c|c|c|c|c|c|c|c|c|}
\hline
Basis & $A_0$ & $A_1$ & $B_0$ & $B_1$ & $c_{00}$ & $c_{01}$ & $c_{10}$ & $c_{11}$ \\
\hline
$ZZ$ &47718&50367&45793&44942&189&7302&7903&250\\\hline
$ZX$ &47117&50726&45467&45831&2735&3826&4075&5061\\\hline
$XZ$ &45985&51051&45509&44441&4077&3643&3806&4317\\\hline
$ZY$ &47775&51018&46149&45415&2579&4382&4650&4545\\\hline
$YZ$ &44564&49626&45739&44157&3382&4155&4414&3505\\\hline
$XX$ &46547&50920&45186&45658&6801&104&148&9083\\\hline
$XY$ &45630&50932&44970&44155&3131&3770&3309&4638\\\hline
$YX$ &44553&49430&45364&45428&3775&3318&2909&4650\\\hline
$YY$ &44499&49666&45718&45152&6586&61&177&8915\\\hline
$\textrm{Dark}$ &418&460&406&440&0&0&0&0\\
\hline
\end{tabular}
\normalsize
\caption{Experimental tomography data for our two-photon experiment. Counts were 1 second in length. A final dark count was taken with the photon source blocked.}
\end{table}

\section{Conclusions}
We have presented a novel method of Bayesian mean estimation using hyperspherical parametrization and an experiment-specific likelihood. This method has allowed us to derive a closed-form BME for the ideal single-qubit and to develop a numerical approach to approximating the BME for a two-qubit experiment using numerical slice sampling. Our approach offers the real world benefit of eliminating the need for preliminary experiments in common two-photon experiments by accounting for qubit loss within the likelihood. Our method is also scalable beyond two-qubit systems. Finally, we illustrated our approach by applying it to the measurement data obtained from a real-world two-photon entangled state. 

\section{Acknowledgement}
We would like to thank Nick Peters, Ryan Bennink, and Robin Blume-Kohout for comments, criticisms, and suggestions regarding this manuscript. This work was supported by the Oak Ridge National Laboratory Postdoctoral Program. This manuscript has been authored by UT-Battelle, LLC, under Contract No. DE-AC05-00OR22725 with the U.S. Department of Energy.

\section{References}
\bibliographystyle{unsrt}
\bibliography{bibliography}


\begin{appendices}
\section{Maximum likelihood estimation with a traditional likelihood}\label{mleAppendix}
Traditional MLE relies on a simple multinomial likelihood that relates the probability of an observation to its quantum probability using Born's Rule,
\begin{equation}P(\mathcal{D}|\alpha)=\prod_{i}\textrm{Tr}\left(E_i\cdot\rho\right)^{c_i}\end{equation}
where $E_i$ are the observable of interest, for instance the Pauli operators $\sigma_z$, $\sigma_x$, $\sigma_y$ or possible Kronecker products of them, $\sigma_z \otimes \sigma_x$, and so forth as dimensionality demands. For instance, the likelihood for a two-photon state with measurement results from a single-basis is
\begin{equation}P(\mathcal{D}|\alpha)=p_{00}^{c_{00}}p_{01}^{c_{01}}p_{10}^{c_{10}}\left(1\!-\!p_{00}\!-\!p_{01}\!-\!p_{10}\right)^{c_{11}}\end{equation}
with $\mathcal{D}=\{c_{00},c_{01},c_{10},c_{11}\}$ and $\alpha=\{p_{00}, p_{01}, p_{10}\}$. The problem with this simple view is that only in an ideal, or possibly perfectly symmetric, experiment does the data set $\mathcal{D}$ truly result from the quantum probabilities alone. Instead the measurement apparatus will add its own bias and inefficiency to the result. Thus, the data set may need to be corrected using experimental constants determined from initial experiments. In this paper we focus on the pathway efficiencies $a_0$, $a_1$, $b_0$ and $b_1$ present in the two-photon apparatus. Given that we know these exactly for a single basis we can correct the data set in the following manner. Identify the smallest efficiency for both Alice and Bob
\begin{equation}a_m=\textrm{min}(a_0,a_1) \quad\quad b_m=\textrm{min}(b_0,b_1)\end{equation}
and use these to correct the counts to
\begin{equation}k_{ij}=\frac{a_m b_m}{a_i b_j}c_{ij}\textrm.\end{equation}
If $a_0$=$a_1$ and $b_0$=$b_1$, it should be apparent that no correction is needed. With the corrected data set $\mathcal{D'}$=$\{k_{00},k_{01},k_{10},k_{11}\}$ maximization of $P(\mathcal{D}'|\alpha)$ can proceed.

\section{Numerically sampling density matrices}\label{ss}
In this Appendix we describe our procedure for sampling density matrices from the true distribution. While other methods of numerical sampling can be used, we have utilized slice sampling \cite{sliceSampling,mackay2003information}. We give a brief description of slice sampling from a single-parameter distribution before describing its extension to density matrix sampling.
\subsection{Slice sampling}
\begin{figure}[b]
\centering
\includegraphics[width=0.55\linewidth]{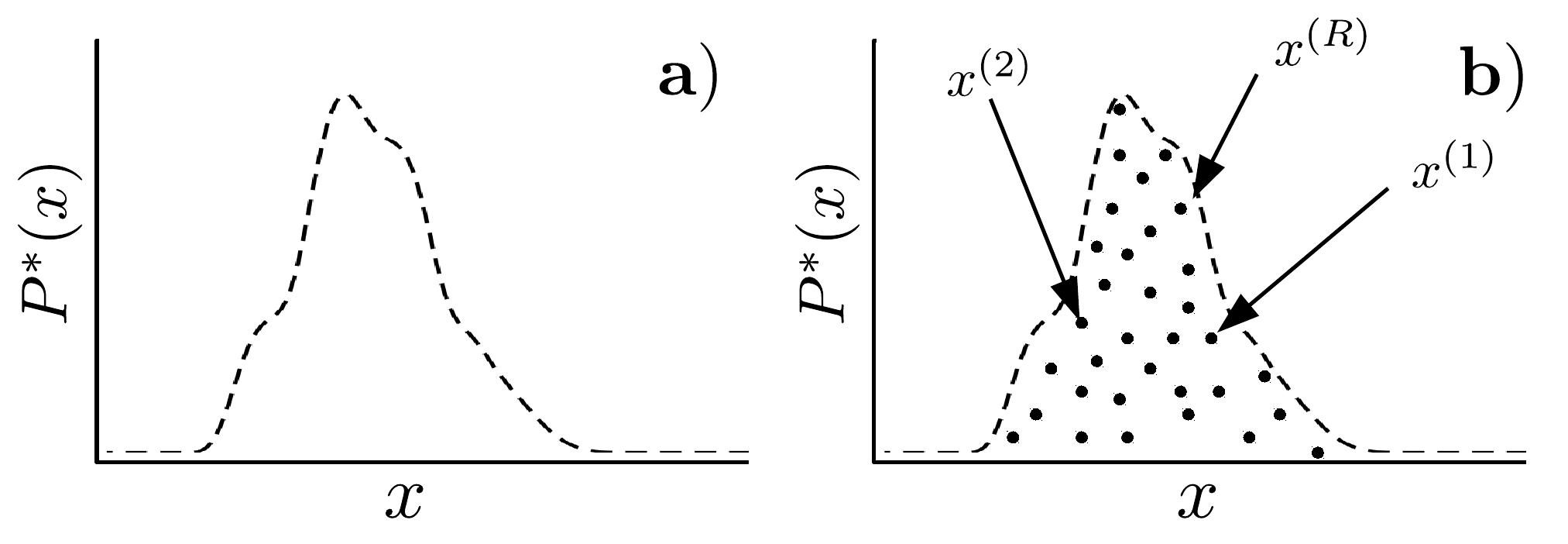}
\caption{If one could fill the space under the curve of an unknown and unnormalized distribution $P^*(x)$ uniformly, each of these points would represent a sample from the true distribution $P(x)$ as the number of points goes to infinity.\label{fill_curve}}
\end{figure}
Slice sampling is an approach to sampling parameter values $x$ from the distribution $P(x)$ given that only the unnormalized distribution $P^*(x)$ is known \cite{sliceSampling, mackay2003information}. This is useful when evaluating $P^*(x)$ everywhere is resource prohibitive--the normalization $Z$=$\int dx P^*(x)$ is unknown. In this case, consider Fig. \ref{fill_curve}a) where the dashed curve represents the likelihood $P^*(x)$ which we do not know everywhere but can evaluate anywhere. Now consider that we can uniformly fill the space under this curve with points as depicted in Fig. \ref{fill_curve}b). If we forget the ``y" coordinate and randomly select one of the $R$ points we would tend to be sampling from the true distribution as $R \rightarrow \infty$. Slice sampling is one method by which to uniformly fill the space under $P^*(x)$ with points.

\begin{figure}[bt]
\centering
\includegraphics[width=0.9\linewidth]{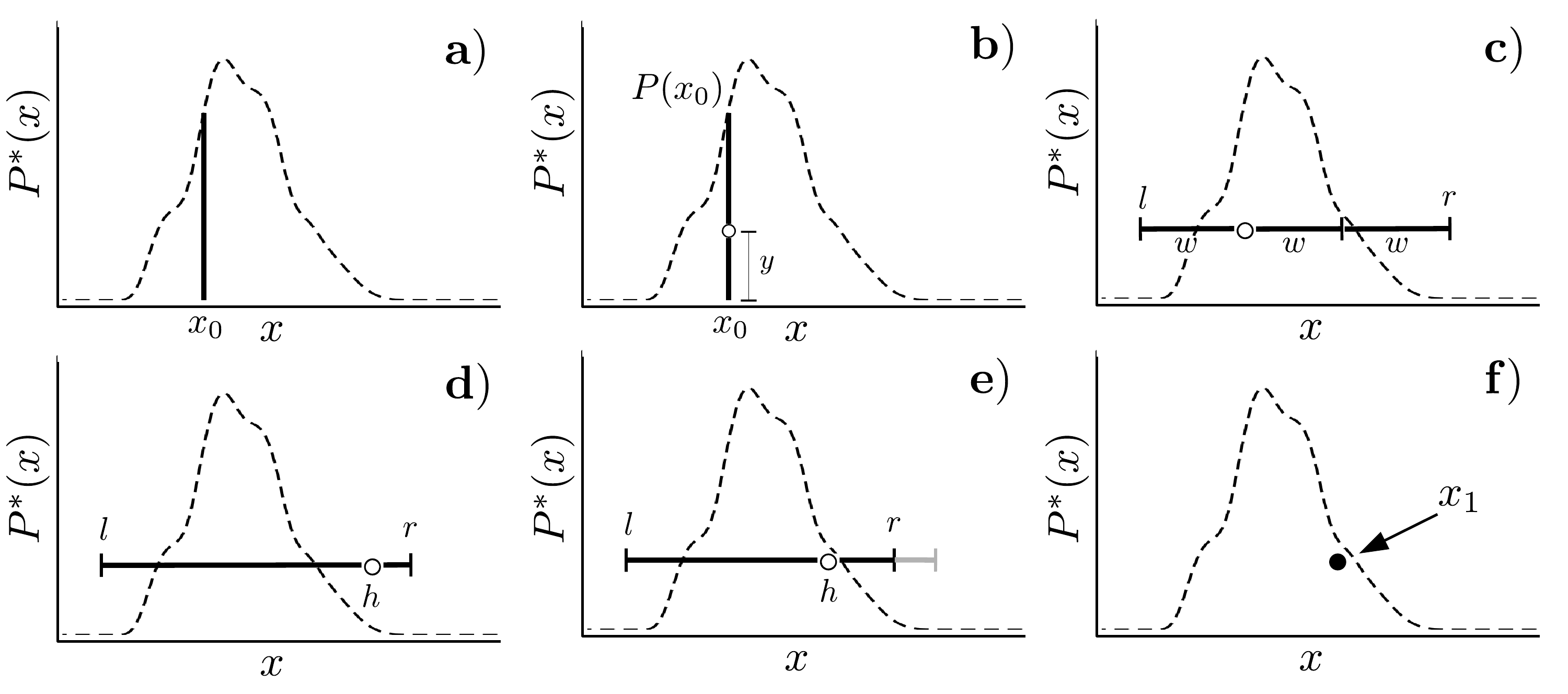}
\caption{This sequence depicts one loop of the the slice sampling algorithm.\label{procedure}}
\end{figure}

There are different slice sampling methods, but we will use that introduced by Neal \cite{sliceSampling,mackay2003information}. Below we evaluate the unnormalized distribution $P^*(x)$ in our sampling algorithm. Referring to Fig. \ref{procedure},
\begin{itemize}
\item[a)] Start at point $x_0$. We assume, for now, this is a random point.
\item[b)] At this location a vertical coordinate $y=q\cdot P^*(x)$, $q\in\left[0,1\right]$, is chosen uniformly. 
\item[c)] From $x_0$ we ``step out" to the left and to the right in steps of $w$ until the left point $l$ and right point $r$ are both outside the curve, $P^*(l)<y$ and $P^*(r)<y$.
\item[d)] We uniformly sample a new point $h$ from the interval $x\in\left[l,r\right]$. If $h$ lies inside the curve, $P^*(h)\geq y$ we accept this as our new point. If $h$ lies outside the curve, $P^*(h)<y$ , we set the new leftmost or rightmost point equal to $h$. Whether $h$ is greater or lesser than $x_0$ reveals which. We do this for our expected unimodal distribution to increase the chance of acceptance for the next sample by eliminating known rejection regions. We reject the first point and shrink the interval.
\item[e)] We select and accept new point $h=x_0$ from the interval, since $P^*(x_1)>y$.
\item[f)] We start over at step a) from point $x_1$.
\end{itemize}

In reality, we use $\ln P^*(x)$ to compare points in the distribution which is more numerically convenient. The first point $x_0$ in step $a)$ has not come from the true distribution. To resolve this, samples can be taken and forgotten until samples are being taken from $P(x)$ with no influence of the starting point $x_0$ present. To evaluate when this point has been reached we utilize multiple independent samplers. Our stopping criteria should ensure that all samplers have converged, and also that enough samples have been taken to sufficiently approximate the distribution $P(x)$. Using multiple samplers has compound utility in that convergence can be assessed and also that more samples can be taken in less time when samplers are run in parallel. Our sampling procedure is to perform a ``burn-in" where each sampler takes $k_0$ samples, the mean $\overline{x}$ and the mean standard deviation $\sigma$ are calculated for each sampler. When the mean of the $i$th sampler $\overline{x}_i$ is separated from each of the other $j$ sampler means $\overline{x}_j$ by less than $m\sigma_j$, $m$  smaller requires closer convergence, we stop the burn in. If this criteria is not met after $k_i$ samples, we forget all but the last sample point, double the number of samples $k_{i+1}=2k_i$, and start over. Once the criteria is met, we repeat the last sampling request for each sampler. The data from this final request is combined as the final set of samples. These samples tend to be from the true distribution $P(x)$ as the convergence parameter $m\rightarrow 0$. 

The estimations we have made in this article have been from distributions assumed to be unimodal. While this is certainly true for the simplest distributions given, it is not conclusive for the more complex distributions we have derived. However our assumption is corroborated by the simulations where multiple samplers are given the same data set, started in random independent locations, and always converge. However, we have not addressed how to approach known multi-modal distributions. In this case, more advanced methods such as sequential Monte Carlo (SMC) \cite{del2006sequential} must be applied to full characterize the posterior distribution. 
\subsection{Density matrix sampling}
The goal of our slice sampling application is to sample density matrices from the true but unknown PD $P(\tau|\mathcal{D})=\frac{1}{Z}P(\mathcal{D}|\tau)P(\tau)$ used in Eq. \ref{PDlast}. To do this we slice sample parameter $x$ in $\tau$ from the conditional unnormalized distribution $P^*\left(x|\tau'\right)$ by keeping all other parameters $\tau'$ fixed, $x\notin \tau'$. The sampling procedure is performed sequentially for $k$, $k\geq n^2-1$, parameters while keeping each new parameter sample point for the next following parameters' conditional likelihoods. Once each parameter has been sampled the new points constitute one sample $\rho^{(r)}$. Beginning with sample point $\rho(\tau^i)$, $\tau^i=\{\tau^i_{1},\tau^i_{2},\cdots,\tau^i_{k}\}$,
\begin{itemize}
\item[1)] sample $t^{i+1}_1$ from $P^*\!\left(\tau_1|\tau^i_{2},\cdots,\tau^i_{k}\right)$
\item[2)] sample $t^{i+1}_2$ from $P^*\!\left(\tau_2|\tau^{i+1}_{1},\cdots,\tau^i_{k}\right)$
\item[j)] sample $t^{i+1}_j$ from $P^*\!\left(\tau_j|\tau^{i+1}_{1},\cdots,\tau^{i+1}_{j-1},\tau^{i}_{j+1},\cdots,\tau^i_{k}\right)$
\item[k)]  sample $t^{i+1}_k$ from $P^*\!\left(\tau_k|\tau^{i+1}_{1},\tau^{i+1}_{2},\cdots,\tau^{i+1}_{k-1}\right)\textrm{.}$
\end{itemize}
The $(i+1)^{\textrm{th}}$ sample is $\rho^{(i+1)}=\rho(\tau^{i+1})$.

Each parameter may have unique ranges, thus slice sampling for each must account for the minimum and maximum parameter values $x\in\{\textrm{min},\textrm{max}\}$. Additionally, some parameters can be cyclic. For instance, angle $\phi\in\{0,2\pi\}$. Extra consideration must be taken for these parameters since their distribution can be centered on a boundary. Hard bounds would not allow this. For these parameters we gave no bounds during the slice sampling algorithm.

\section{Maximization using gradient ascent}\label{searchAppendix}
\begin{figure}[tb]
\centering
\includegraphics[width=0.9\linewidth]{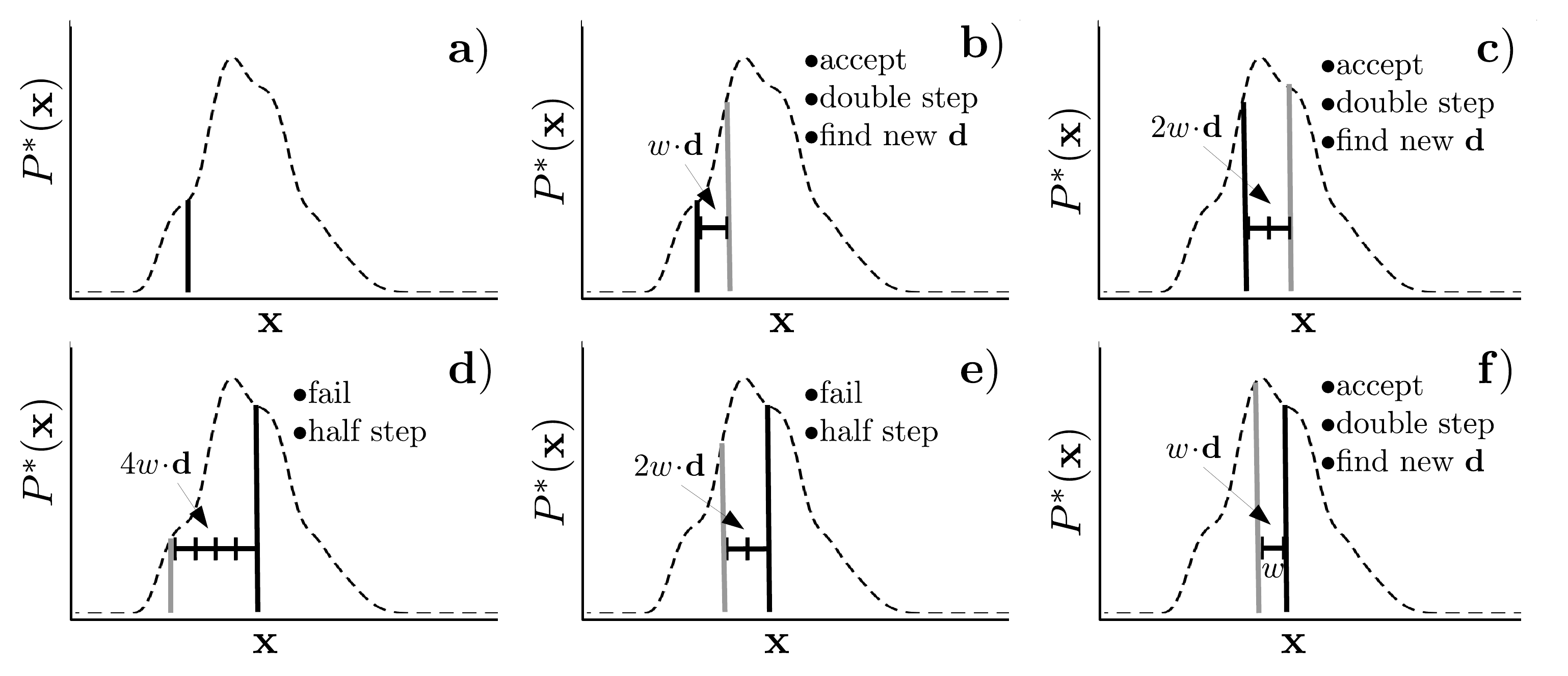}
\caption{This sequence depicts a few iterations of the gradient ascent method. The bottom axis represents an idealized axis along the line of highest gradient ascent direction within the multi-dimensional space.\label{search}}
\end{figure}
To locate a local maximum from the likelihood we use an gradient ascent method as shown in Fig. \ref{search}. This methods auto-tunes the step size to make both large and small steps when appropriate.
\begin{itemize}
\item[a)] Starting from the current multi-dimensional point $\textbf{x}$ the gradient $\nabla P^*(\textbf{x})$ is determined using the finite difference method. The direction of ascent $\textbf{d}$ is found by normalizing the gradient with the $\ell^2$ norm
\begin{align}\textbf{d}&=\frac{\nabla P^*(\textbf{x})}{|\nabla P^*(\textbf{x})|}\nonumber\\
|\nabla P^*(\textbf{x})|&=\sqrt{\sum_i \left(\frac{\partial P^*(\textbf{x})}{\partial x_i}\right)^2}\nonumber\end{align}
\item[b)] A step of size $w$ is made in the direction of increase, $\textbf{x}\rightarrow \textbf{x}+w\cdot\textbf{d}$. This point is accepted since $P^*(\textbf{x}+w\cdot\textbf{d}) \geq P^*(\textbf{x})$. The step size is doubled.
\item[c)] The direction of ascent $\textbf{d}$ is found. A step of $2w$ is made in the direction of increase, $\textbf{x}\rightarrow \textbf{x}+2w\cdot\textbf{d}$. This point is accepted since $P^*(\textbf{x}+2w\cdot\textbf{d}) \geq P^*(\textbf{x})$. The step size is doubled.
\item[d)] The direction of ascent $\textbf{d}$ is found. A step of $4w$ is made in the direction of increase, $\textbf{x}\rightarrow \textbf{x}+4w\cdot\textbf{d}$. This point is not accepted since $P^*(\textbf{x}+4w\cdot\textbf{d}) < P^*(\textbf{x})$. The step size is halved.
\item[e)] A step of $2w$ is made in the direction of increase, $\textbf{x}\rightarrow \textbf{x}+2w\cdot\textbf{d}$. This point is not accepted since $P^*(\textbf{x}+2w\cdot\textbf{d}) < P^*(\textbf{x})$. The step size is halved.
\item[f)] A step of $w$ is made in the direction of increase, $\textbf{x}\rightarrow \textbf{x}+w\cdot\textbf{d}$. This point is accepted since $P^*(\textbf{x}+w\cdot\textbf{d}) \geq P^*(\textbf{x})$. The step size is doubled.
\end{itemize}
In this way, the local maximum is ultimately approached.
\end{appendices}
\end{document}